\documentclass[twocolumn]{aastex701}

\accepted{September 15, 2026}

\graphicspath{{./}{}}

\defcitealias{Cardamone2009}{C09}
\defcitealias{Brunker2020}{B20}
\defcitealias{Kimsey-Miller2024}{KM24}
\defcitealias{Salzer2023}{SFACT1}
\defcitealias{Sieben2023}{SFACT2}
\defcitealias{Carr2023}{SFACT3}
\usepackage[normalem]{ulem} 
\usepackage{xcolor }
\begin{document}

\title{Searching for Green Pea Galaxies in the SFACT Survey}

\author[orcid=0009-0007-4371-9882,sname=Baker]{Kristin N. Baker}
\affiliation{Department of Astronomy, Indiana University, 727 East Third Street, Bloomington, IN 47405, USA}
\email[show]{knb1@iu.edu}  

\author[orcid=0000-0001-8483-603X]{John J. Salzer}
\affiliation{Department of Astronomy, Indiana University, 727 East Third Street, Bloomington, IN 47405, USA}
\email{knb1@iu.edu}  

\author[orcid=0009-0005-6705-7073]{Brooke Kimsey-Miller}
\affiliation{Department of Astronomy, Indiana University, 727 East Third Street, Bloomington, IN 47405, USA}
\email{knb1@iu.edu}

\author[orcid=0000-0001-6776-2550]{Samantha W. Brunker}
\affiliation{NorthWest Arkansas Community College, Bentonville, AR 72712, USA}
\email{}

\author[orcid=0000-0002-4876-5382]{David J. Carr}
\affiliation{Department of Astronomy, Indiana University, 727 East Third Street, Bloomington, IN 47405, USA}
\email{}

\author[orcid=0000-0002-5513-4773]{Jennifer Sieben}
\affiliation{Department of Astronomy, Indiana University, 727 East Third Street, Bloomington, IN 47405, USA}
\email{}

\author[orcid=0000-0001-9165-8905]{Steven Janowiecki}
\affiliation{University of Texas, Hobby–Eberly Telescope, McDonald Observatory, TX 79734, USA}
\email{}

\author[orcid=0000-0003-2342-7501]{Samir Salim}
\affiliation{Department of Astronomy, Indiana University, 727 East Third Street, Bloomington, IN 47405, USA}
\email{}

\begin{abstract}

Green Pea galaxies (GPs) are a class of rare, extreme star-forming galaxies whose physical properties closely resemble high-redshift Ly-$\alpha$ emitting galaxies. Their high equivalent width [\ion{O}{3}] emission lines make them readily identifiable in surveys of emission-line galaxies like the on-going deep, narrowband Star Formation Across Cosmic Times (SFACT) survey. In this study, we use broadband gri colors and [\ion{O}{3}] equivalent widths derived from narrowband images to select a new sample of $\sim$120 GP candidates from a catalog of [\ion{O}{3}]-detected galaxies identified in the SFACT survey. All SFACT GP candidates have spectroscopically confirmed redshift in the range z~=~0.31~--~0.51. The median apparent magnitude of our sample is g~=~22.9. We compare the properties of our GP candidate sample to other samples of known GPs. We find that the excellent depth of SFACT allows us to detect GP candidates to lower luminosities than these comparison samples. This is the first catalog of SFACT GP candidates; additional candidates will be added as the spectroscopic follow-up of the SFACT survey is completed. 

\end{abstract}

\keywords{\uat{Emission line galaxies}{459} --- \uat{Starburst galaxies}{1570} --- \uat{Star formation}{1569} --- \uat{Galaxy evolution}{594}} 


\section{Introduction} \label{sec:intro}
Since their discovery by citizen scientists in the Galaxy Zoo project \citep{Lintott2008}, Green Pea galaxies (GPs) have attracted significant attention due to their extreme physical properties. GPs are a class of star-forming galaxies with strong [\ion{O}{3}] $\lambda$5007 emission (equivalent widths up to $\sim$2000~$\text{\AA}$; \citealt{Cardamone2009}, hereafter \citetalias{Cardamone2009}), high star formation rates (SFRs) and specific SFRs (sSFRs; $\sim$10$^{-9}$~--~10$^{-7}$~yr$^{-1}$; \citetalias{Cardamone2009}; \citealt{Izotov2011}; \citealt{Brunker2020}), low gas-phase metal abundances (12~+~log(O/H)~$\sim$~7.5~--~8.5; \citealt{Amor2010}; \citealt{Izotov2011}), and compact geometries (r$_{h}$ $\lesssim$ 5 kpc; \citetalias{Cardamone2009}). Compared to typical star-forming galaxies (SFGs) with the same masses, GPs tend to lie below the main locus of the  mass-metallicity relation (e.g., \citealt{Amor2010}; \citealt{Brunker2020}) and above the main locus of the star forming main sequence (e.g., \citetalias{Cardamone2009}; \citealt{Izotov2011}; \citealt{Brunker2020}; \citealt{Liu2022}; \citealt{Ding2023}; \citealt{Jeong2025}). 

GPs are named such because of their appearance in Sloan Digital Sky Survey \citep[SDSS][]{York2000} \textit{gri} three-color images. At their discovery redshift range of z~=~0.14~--~0.35, their high equivalent width $\lambda$5007 line is redshifted into the SDSS \textit{r} filter. When the \textit{r} filter is mapped to green in the three-color images, the GPs appear compact and green. The extreme properties of GPs make it possible to detect them in multiple ways. \citetalias{Cardamone2009} cataloged the first sample of GPs in SDSS using broadband color selection techniques. Their criteria work well for the redshift range z~=~0.14~--~0.35, where the [\ion{O}{3}] $\lambda$5007 line falls in the SDSS \textit{r} filter. Many other studies have relied on color criteria to catalog other samples of GPs (\citealt{Liu2022}; \citealt{Jeong2025}; \citealt{Samonski2025}). However, GP-like galaxies that lie to higher and lower redshifts are missed by the criteria established in \citetalias{Cardamone2009} (see Section \ref{sec:SFACT_GPs}). Because of their strong [\ion{O}{3}]  $\lambda$5007 lines, GPs can be detected over broader redshift ranges in samples using emission-line selection methods \citep{Izotov2011,Jiang2019,Brunker2020,Ding2023,Kimsey-Miller2024}.

The discovery of GPs has opened a broader inquiry into a general population of compact, high excitation systems with large EW [\ion{O}{3}] and H$\alpha$ emission lines. These so called extreme emission line galaxies (EELGs) overlap with GPs in parameter space (\citealt{Amor2015}), but span a wider range of redshifts (z~$\sim$~0~$-$~9; \citealt{vanderWel2011}; \citealt{Maseda2014}; \citealt{Amor2015}; \citealt{Tran2020}; \citealt{perez-montero2021}; \citealt{JPLUS2022}; \citealt{Davis2024}; \citealt{Boyett2024}; \citealt{JPAS2026}; \citealt{Llerena2026}) and stellar masses (log(M$_{*}$/M$_{\odot}$)~$\sim$~6~$-$~11; \citealt{vanderWel2011}; \citealt{Maseda2014}; \citealt{Amor2015}; \citealt{JPLUS2022}; \citealt{Boyett2024}; \citealt{Llerena2026}; \citealt{Bonatto2026}; see Figure 4 from \citealt{Hirschauer2022}). Samples of EELGs find similarly elevated sSFRs ($\sim$10$^{-9}$~--~10$^{-7}$~yr$^{-1}$; \citealt{vanderWel2011}; \citealt{Maseda2014}; \citealt{Amor2015}; \citealt{Tran2020}; \citealt{JPLUS2022}; \citealt{Boyett2024}; \citealt{Llerena2026}; \citealt{JPAS2026}) and low gas-phase metal abundances (12~+~log(O/H)~$\sim$~7.5~--~8.5; \citealt{Amor2015}; \citealt{Boyett2024}) compared to GPs. EELGs may represent a broad general population of compact galaxies undergoing an intense burst of star formation (\citealt{vanderWel2011}; \citealt{Amor2015}; \citealt{Boyett2024}) with varying evolutionary pathways. The goal of the current study is to assemble a sample of extreme star-forming galaxies that resemble the \citetalias{Cardamone2009} GP sample as closely as possible.

The extreme properties of GPs and other EELGs mirror higher redshift galaxies present during the Epoch of Reionization (EoR). During this era, the universe underwent a major phase change from a fully neutral intergalactic medium (IGM) to a mostly ionized IGM. Reionization is mostly concluded by z $\sim$ 6 \citep[][]{Fan2006,McGreer2015}, with SFGs \citep{Robertson2010}, active galactic nuclei \citep[AGNs][]{Haiman1998,Madau2015}, and quasars \citep{Madau2004} being candidates for sources of ionizing photons. SFGs are thought to be the main contributors to reionization. However, the large cold gas content in SFGs typically prevents ionizing photons produced by hot stars from escaping into the IGM. Due to their extreme nature, many GPs are confirmed to have escaping Ly$\alpha$ emission \citep{Henry2015,Yang2016,Yang2017,Kim2021,Amor2024}. GPs are among the only known galaxies to have confirmed escaping Lyman continuum radiation \citep{Izotov2016Natur, Izotov2016MNRAS, Izotov2017, Izotov2018,Jaskot2017,Jaskot2025}. While GPs \citep{Rhoads2023} and GP-like EELGs \citep{Katz2023,Trump2023,Laseter2025} have been found at high redshifts, low-redshift GPs and GP-like EELGs remain ideal local laboratories for understanding physical processes happening in the high-redshift universe, as the proximity of  these low-redshift analogs enables high signal-to-noise optical spectroscopy across a well-characterized spectral region and allows for multi-wavelength observations to further contextualize observations \citep{Brinchmann2023}.

In the current study, we present a new, preliminary sample of GP candidates detected in the Star Formation Across Cosmic Time survey \citep[SFACT;][]{Salzer2023,Sieben2023,Carr2023}, a narrowband imaging survey. We select GP candidates from an emission-line detected sample in three narrow redshift windows using broadband color and $\lambda$5007 equivalent width criteria. We then compare properties of the SFACT GP candidates to other samples of known GPs, including the original \citetalias{Cardamone2009} GP sample. The use of SFACT to select a sample of GPs offers advantages compared to previous studies.  By pre-selecting objects with strong emission, we aim to construct a GP candidate sample with a well-characterized line-flux limit.

In Section \ref{sec:data}, we outline the overall design and status of the SFACT survey. In Section \ref{sec:O3_det}, we describe our methods for using the narrowband photometry to derive photometric equivalent widths of the [\ion{O}{3}] $\lambda$5007 line. In Section \ref{sec:SFACT_GPs}, we present the selection criteria used to identify SFACT GP candidates. In Section \ref{sec:prop}, we discuss and compare the properties of the SFACT GP candidates with other samples of GPs. We assume a standard $\Lambda$CDM cosmology with $H_{0}$~=~70~kms$^{-1}$Mpc$^{-1}$, $\Omega_{m}$~=~0.27, and $\Omega_{\Lambda}$~=~0.73 for our calculations. 

\section{The Star Formation Across Cosmic Time Survey} \label{sec:data}
The GP candidates identified in this work were discovered as part of the SFACT survey \citep{Salzer2023,Sieben2023,Carr2023}. SFACT is an ongoing survey carried out on the WIYN~3.5~m telescope that uses narrowband (NB) imaging and spectroscopy to detect and to verify a large number of emission-line sources across a wide range of redshifts. The goal of SFACT is to produce a statistically complete catalog of emission-line galaxies (ELGs) useful for a broad range of science applications, including the study of Green Peas. \citet[][hereafter \citetalias{Salzer2023}]{Salzer2023} presents an overview and the results from the survey’s pilot-study fields. 

\begin{figure}[t!]
\centering
\includegraphics[width=\linewidth]{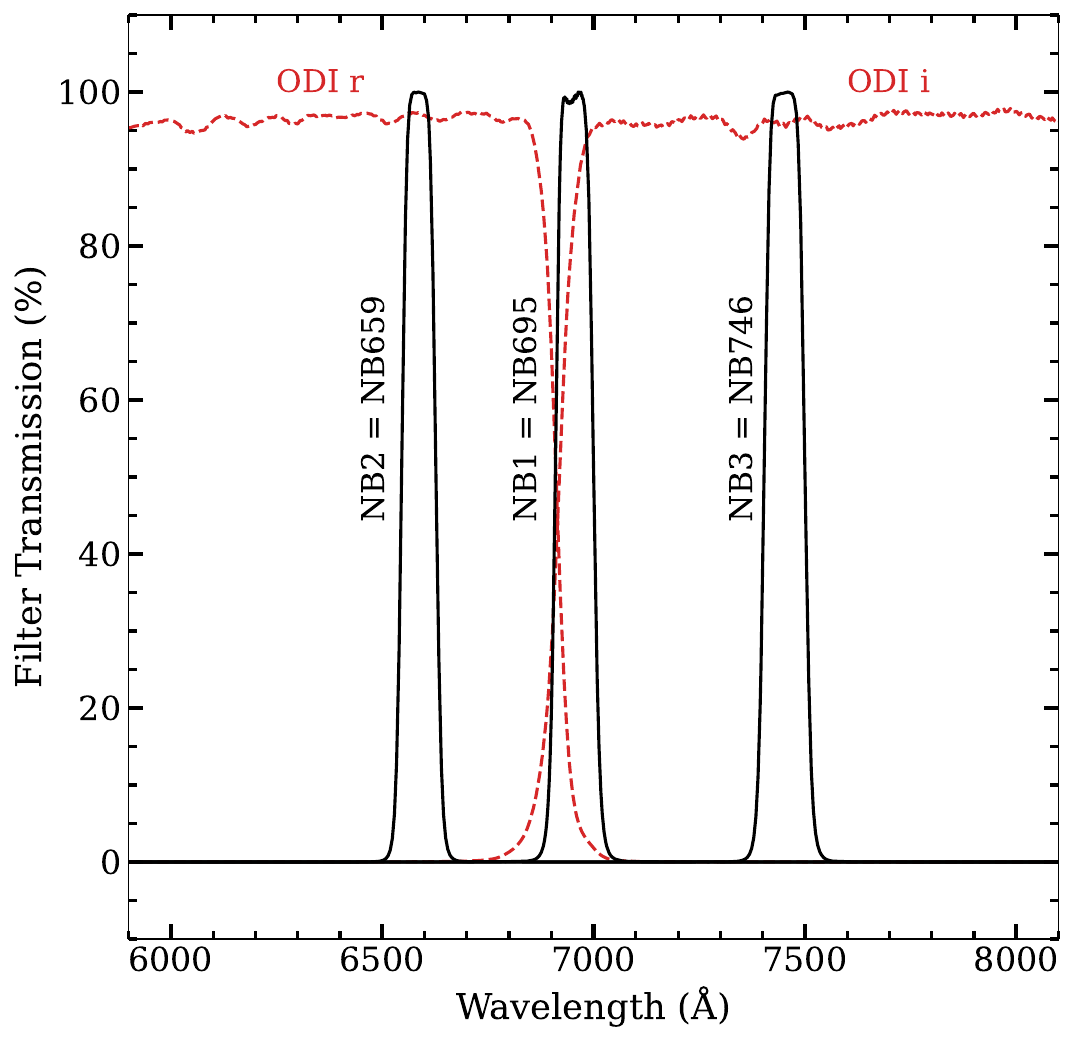}  
\caption{The transmission curves of three narrowband filters used in the imaging portion of the SFACT survey. The filters denoted as NB659, NB695, and NB746 are centered at 6590~$\text{\AA}$, 6950~$\text{\AA}$, and 7460~$\text{\AA}$, respectively. The red lines show part of the transmission curves for the ODI \textit{r} and \textit{i} filters.}
\label{fig:trans_curves}
\end{figure}

SFACT imaging data are obtained with the One Degree Imager \citep[ODI;][]{Harbeck2010} camera on the WIYN~3.5~m Telescope. ODI has a field of view of 40$\arcmin$~$\times$~48$\arcmin$ and a native pixel scale of 0$\farcs$11~pixel$^{-1}$. All survey fields are observed through \textit{gri} broadband (BB) filters and three NB filters. All observations are acquired using a nine-point dither pattern to eliminate chip gaps present in the ODI focal plane. Exposure times of 120 s and 600 s are used for each dither pattern image for the BB images and NB images, respectively. \citet[][hereafter \citetalias{Sieben2023}]{Sieben2023} further details the data acquisition and image processing. 

The NB data were taken with custom filters designed for the survey. The filters are centered on 6590~$\text{\AA}$ (referred to as NB2 or NB659 hereafter), 6950~$\text{\AA}$ (NB1 or NB695), and 7460~$\text{\AA}$ (NB3 or NB746), each with a width of $\sim$90~$\text{\AA}$. The transmission curves of the three NB filters are shown in Figure \ref{fig:trans_curves}. The NB filters fall within the wavelength coverage of the \textit{r} or \textit{i} BB filters. 

Objects are found automatically in each image set using scripts that calls \texttt{DAOFIND} \citep{stetson1987}. SFACT identifies ELG candidates by computing the difference, or $\Delta$m, in instrumental magnitudes measured from the NB images and scaled BB images (\citetalias{Sieben2023}). The $\Delta m$ value is a measure of the excess in NB flux an object has compared to the continuum flux. Observed objects are considered ELG candidates if they exhibit a statistically significant excess (greater than 5$\sigma$) of NB flux that is 0.4 mag brighter than the continuum flux, or $\Delta m\geq-0.4 \text{ mag}$. SFACT ELG candidates that demonstrate this NB excess have median g-band magnitude of $g\sim23$, with the faintest ELG candidates having $g\sim26$ (\citetalias{Salzer2023}). SFACT is complete to emission-line fluxes of $\sim$$1.9 \times10^{-16} \text{ erg s}^{-1}\text{ cm}^{-2}$ (\citetalias{Salzer2023}). See \citetalias{Sieben2023} for more details on SFACT candidate selection.

Follow-up spectroscopic data are obtained for all SFACT candidates using the Hydra multi-fiber positioner \citep{Barden1994, Barden1995} and Bench spectrograph on the WIYN~3.5~m telescope. Hydra has a field of view of 1$\fdg$0 in diameter, fully encompassing the areal coverage of the ODI camera. SFACT utilizes the Hydra red fibers, subtending 2$\arcsec$ on the sky, to obtain spectra for 60-65 targets per configuration. The Bench spectrograph is configured with the 600 @ 10.1 bench grating which delivers a dispersion of 1.41~$\text{\AA}$~pixel$^{-1}$ and a spectral resolution of 3.35~$\text{\AA}$, corresponding to the FWHM of the line spread function. With this configuration, SFACT spectra have an observed wavelength range of $\lambda\lambda$4760~--~7580~$\text{\AA}$, which covers the wavelength ranges of all three NB filters. Each configuration is observed for three exposures of 30 minutes each to mitigate effects of cosmic rays and artifacts. The spectroscopic data are processed and measured as described in \citet[][hereafter \citetalias{Carr2023}]{Carr2023}. During these steps, all spectra are visually inspected. Any SFACT candidates found to have broad lines (e.g., Seyfert 1s and QSOs) are labeled as AGNs and are not considered further as possible GP candidates. See \citetalias{Carr2023} for more details.  

\begin{figure}[t!]
\centering
\includegraphics[width=\linewidth]{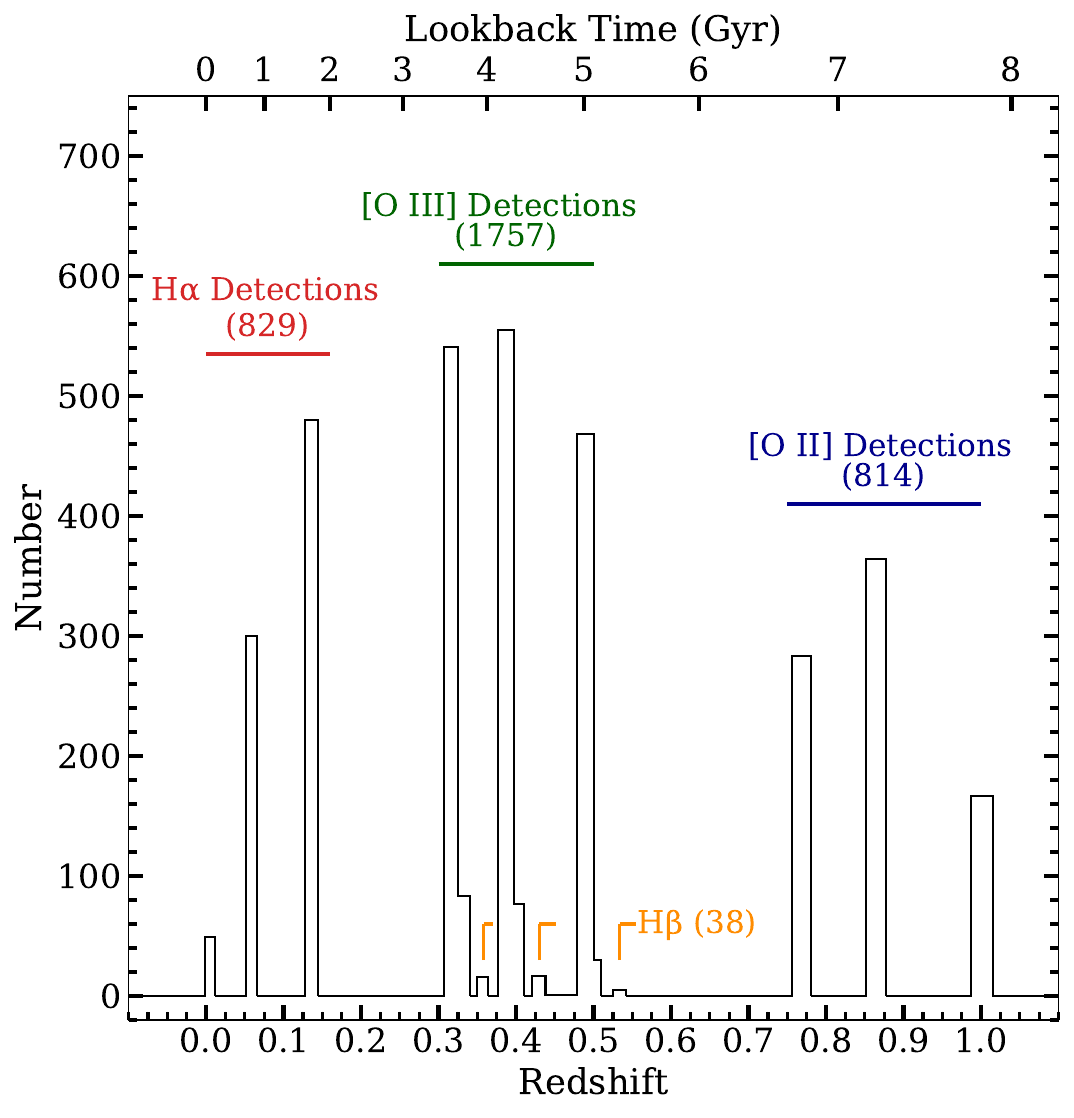} 
\caption{Histogram of the redshift distribution of ELGs detected by H$\alpha$, [\ion{O}{3}], H$\beta$, or [\ion{O}{2}] within the SFACT survey. The use of NB imaging leads to the detection of ELGS within discrete redshift widows across a wide redshift range (z $\approx$ 0 -- 1). }
\label{fig:sfact_reds}
\end{figure}

SFACT ELGs are detected primarily via H$\alpha$, [\ion{O}{3}] $\lambda$5007, or [\ion{O}{2}] $\lambda$3727 in one of the NB filters. The selection threshold of excess NB flux employed by SFACT corresponds to rest-frame equivalent width selection limits of $\sim$36--39~$\text{\AA}$ for H$\alpha$ detections, $\sim$27--30~$\text{\AA}$ for [\ion{O}{3}] detections, and $\sim$20--22~$\text{\AA}$ for [\ion{O}{2}] detections (\citetalias{Sieben2023}). The use of NB imaging in this way leads to ELGs being detected across a wide redshift range (z~$\approx$~0~--~1) within discrete redshift widows. This is illustrated in the redshift distribution of the current SFACT catalog presented in Figure \ref{fig:sfact_reds}. The width of each bin corresponds to the observed width of each redshift window. The redshift windows shown break down into three principal groups: H$\alpha$ detections at z~$\approx$~0~--~0.15, [\ion{O}{3}] $\lambda$$\lambda$4959,5007 detections at z~$\approx$~0.31~--~0.51, and [\ion{O}{2}] $\lambda$3727 detections at z~$\approx$~0.76~--~1.01. See \citetalias{Salzer2023} for more details.

The original survey papers (\citetalias{Salzer2023}, \citetalias{Sieben2023}, and \citetalias{Carr2023}) discuss three pilot study fields with 415 spectroscopically confirmed ELGs and a sky coverage of 1.59~deg$^{2}$. Because SFACT is an ongoing survey, the current study utilizes a larger data set obtained from observations made through Fall 2024. The full SFACT catalog used in this work consists of 3,658 spectroscopically confirmed ELGs detected in 41 fields spanning all redshift windows and covering 21.73~deg$^{2}$. These fields are scattered across the sky with 24 fields in the Fall sky and 17 in the Spring sky.

From this parent sample, we draw our SFACT GP candidate sample from 1,757 galaxies detected by [\ion{O}{3}] $\lambda$$\lambda$4959,5007. We apply a photometric error cut of $\sigma_{r}$~$\leq$~0.1 in the \textit{r} band to exclude detections with unreliable photometry. We note that this requirement decreases the magnitude limit of the sample, resulting in the faintest detections having $r\sim23$ compared to $r\sim25$ for the full sample. However, the selection criteria developed in this work require accurate photometry to isolate GPs in the overall sample (see Section \ref{sec:SFACT_GPs}). The final sample from which we select SFACT GP candidates consists of 1,260 spectroscopically confirmed [\ion{O}{3}]-detected galaxies with $\sigma_{r}$~$\leq$~0.1, covering 21.73~deg$^{2}$. 

The ``first look" spectroscopic portion of SFACT survey is currently in progress. Spectroscopic observations for the SFACT catalog utilized in this work are roughly three-quarters complete for the 41 fields surveyed. Because the SFACT survey is ongoing, the GP candidate catalog presented in this work is preliminary, and we expect the list of SFACT GP candidates to grow as more spectroscopic data are obtained. Because of this, we are currently unable to properly determine the completeness or purity of the sample. However, we plan to publish a supplementary SFACT GP candidate list once SFACT is finished, at which time we will be able to quantify the depth and completeness of the SFACT GP sample. We are also in the process of obtaining higher quality spectra with a wider wavelength coverage which will allow us to further assess the nature of the SFACT GP candidate sample.

\section{SFACT [\ion{O}{3}]-Detected Galaxies} \label{sec:O3_det}
 
A defining characteristic of GP galaxies is their strong [\ion{O}{3}] emission lines, making them identifiable in samples of ELGs like SFACT. This work focuses solely on the SFACT [\ion{O}{3}] detections to construct a new, deep sample of GPs. The use of three NB filters leads to three redshift windows over which galaxies can be detected via their [\ion{O}{3}] lines: (1) z~=~0.31~--~0.34 for NB659 [\ion{O}{3}] detections; (2) z~=~0.38~--~0.41 for NB695 [\ion{O}{3}] detections; and (3) z~=~0.48~--~0.51 for NB746 [\ion{O}{3}] detections.  While GPs are typically selected by the strength of their $\lambda$5007 emission, we also include objects that were detected solely via $\lambda$4959 lines. We note that sources with sufficient $\lambda$4959 emission to be detected in SFACT will have $\sim$2.9 $\times$ stronger $\lambda$5007 emission and are likely to be some of the most extreme sources detected. 

\begin{figure}[t!]
\centering
\includegraphics[width=\linewidth]{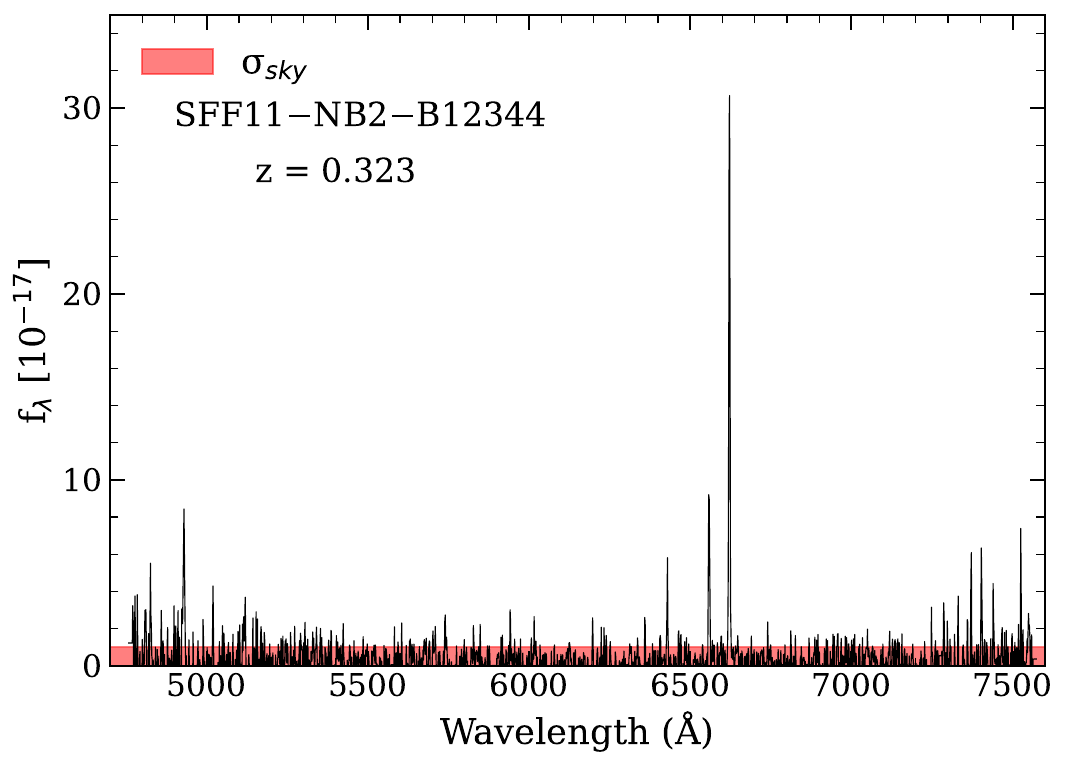} 
\caption{Hydra spectrum of an SFACT source, SFF11-NB2-B12344, detected by the [\ion{O}{3}] lines in the NB659 filter. The red shaded region denotes the 1$\sigma$ uncertainty of 1~$\times$~10$^{-17}$~erg~s$^{-1}$~cm$^{-2}$~$\text{\AA}$$^{-1}$ in the subtraction of the composite sky continuum. The measured continuum level after the sky subtraction lies below the 1$\sigma$ uncertainty. The \textit{r}-band and \textit{i}-band apparent magnitudes for this source are 20.881 and 21.584, respectively.}
\label{fig:sfact_spec_cont}
\end{figure}

\begin{figure}[t!]
\centering
\includegraphics[width=\linewidth]{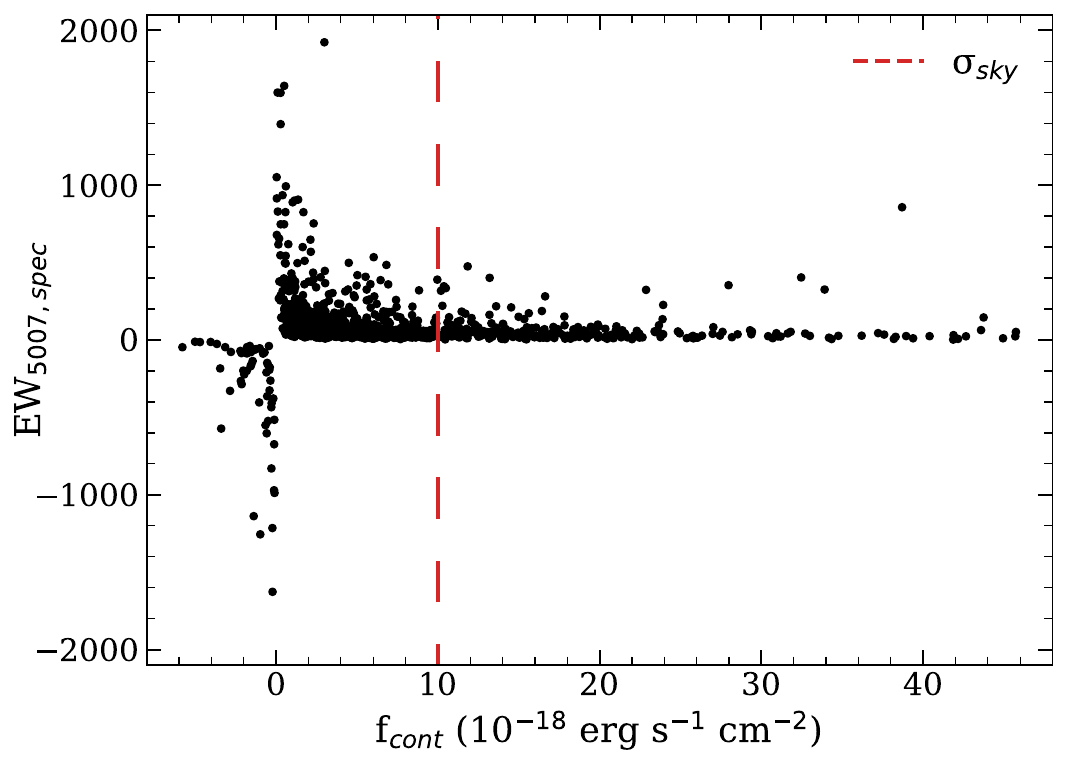} \caption{Observed spectroscopic [\ion{O}{3}] $\lambda$5007 EW plotted against the measured continuum flux for 1,260 SFACT [\ion{O}{3}]-detected galaxies with $\sigma_{r} \leq0.1$.  The continuum flux is measured automatically for each SFACT spectrum using the ALFA software \citep{wesson2016} described in \citetalias{Carr2023}.  The red dashed line delimits the 1$\sigma$ uncertainty in the sky subtraction. The measured EWs become increasingly unreliable below this flux level. }
\label{fig:cont_flux}
\end{figure}

A defining criterion used in some previous studies that defined GP samples \citep{Jiang2019,Ding2023} is the [\ion{O}{3}] $\lambda$5007 equivalent width (EW) measured from spectra. Due to limitations of the Hydra spectra, we cannot rely on the measured spectroscopic EWs to construct our sample. While SFACT detections have strong emission, most sources have faint underlying continua. In many cases, the uncertainty in the subtraction of the composite sky spectrum, which has a characteristic value of 1.0 $\times$ 10$^{-17}$ erg~s$^{-1}$~cm$^{-2}$~$\text{\AA}$$^{-1}$,   results in exceedingly faint or even negative continua. This is illustrated in the reduced spectrum of an  [\ion{O}{3}] detection shown in Figure \ref{fig:sfact_spec_cont}. The continuum level after the sky spectrum is subtracted lies below the 1$\sigma$ uncertainty in the sky subtraction. The effect of this is evident in the equivalent widths measured from the Hydra spectra. Figure \ref{fig:cont_flux} plots the spectroscopically measured EW of the $\lambda$5007 line (EW$_{5007,spec}$) versus the measured continuum flux used to calculate the EW$_{5007,spec}$ for all SFACT [\ion{O}{3}] detections. The majority of the SFACT detections have continuum flux levels below the 1$\sigma$ sky subtraction uncertainty indicated by the dashed line. These sources have measured EW$_{5007,spec}$ that are, in some cases, unphysically large or negative. Hence, the use of measured EW$_{5007,spec}$ for GP selection is not possible. 

Since the spectroscopic EW$_{5007,spec}$ measurements are not universally reliable, we instead use the fluxes from SFACT NB images to construct a ``photometric EW" for the $\lambda$5007 line (EW$_{5007,phot}$) for each SFACT [\ion{O}{3}] detection. This photometric EW will serve as a surrogate for the more commonly used spectroscopic EW (EW$_{5007,spec}$) for the purpose of identifying GPs in SFACT. We note that several previous NB studies have utilized a combination of NB and BB photometry to calculate photometric EWs (\citealt{Waller1990,Sobral2009,Sobral2012,Stroe2015,Nandi2023}; \citealt{JPLUS2022,miniJPAS2022,JPAS2026}).

\begin{figure*}[t!]
\centering
\includegraphics[width=\linewidth]{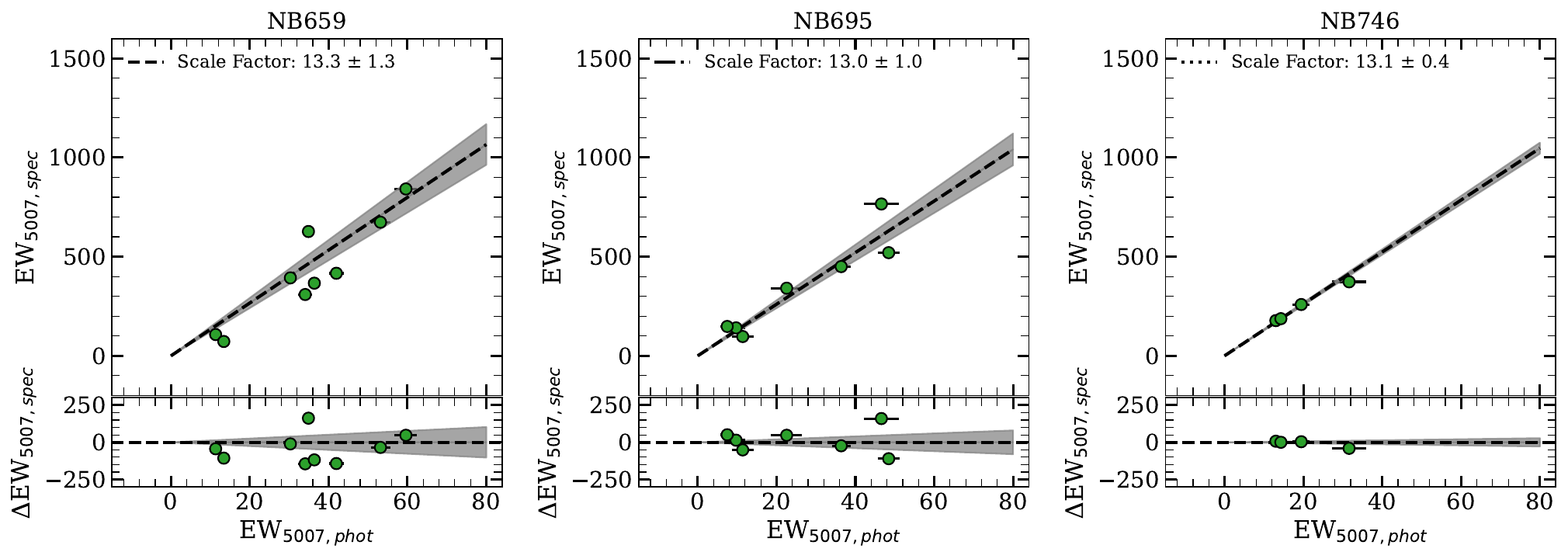} 
\caption{\textbf{\textit{Top:}} The observed EW$_{5007,spec}$ measured from the HET spectra plotted against the photometrically derived EW$_{5007,phot}$ values for the NB659 (left), NB695 (middle), and NB746 (right) filters. The black dashed lines are the linear least squares fits to the data for each filter. The slopes of these lines are used to scale the EW$_{5007,phot}$ values. The grey shaded regions represent the 1$\sigma$ errors on the scale factors. The best fit lines are anchored at EW$_{5007}$ of zero for both photometric and spectroscopically determined EW$_{5007}$. \textit{Bottom:} Residual plots for the linear least squares fits.}
\label{fig:HET_scale}
 \end{figure*}
 
In order to derive the EW$_{5007,phot}$, we first find the photometric equivalent width of the flux in the NB filter (EW$_{NB}$) by dividing the calibrated, continuum subtracted, NB flux by the continuum flux. The NB flux for each detected SFACT source is cataloged in units of  erg~s$^{-1}$~cm$^{-2}$ (see \citetalias{Sieben2023} for details).  We use the average flux of the \textit{r} and \textit{i} filters as an approximation for the continuum flux for [\ion{O}{3}] detections in the NB659 and NB695 filters, and the flux in the \textit{i} filter for NB746 [\ion{O}{3}] detections. This mirrors the use of these BB images for continuum subtraction of the NB images as described in \citetalias{Sieben2023}. These continuum fluxes are derived by converting the calibrated \textit{r} and \textit{i} magnitudes into flux densities with units of erg~s$^{-1}$~cm$^{-2}$~$\text{\AA}$$^{-1}$ using the standard conversion equation \citep[e.g.,][]{massey1988}.

In cases where [\ion{O}{3}] $\lambda$5007 is the only line in NB filter, we derive EW$_{5007,phot}$ by applying a bandpass correction to the EW$_{NB}$ to account for the wavelength-dependent transmission of the filter, effectively normalizing the line flux to its value at the central wavelength of the filter:
\begin{equation}
EW_{5007,phot}~=~\frac{\mathrm{EW_{NB}}}{\mathrm{T_{NB}}(\lambda_{5007,obs})},
\end{equation}
\noindent where T$_{NB}$($\lambda_{5007,obs}$) is the normalized filter transmission at the observed (redshifted) wavelength of $\lambda$5007.  For galaxies where both [\ion{O}{3}] lines are in the NB filter we assume $\lambda$5007 would have 2.9 times the flux of $\lambda$4959 \citep{Osterbrock2006}, and our bandpass correction to derive EW$_{5007,phot}$ becomes:
\begin{equation}
EW_{5007,phot}~=~\frac{\mathrm{EW_{NB}}}{\mathrm{T_{NB}}(\lambda_{5007,obs}) + 0.34\cdot (\mathrm{T_{NB}}(\lambda_{4959,obs}))},
\end{equation}
\noindent where T$_{NB}$($\lambda_{4959,obs}$) is the normalized filter transmission at the observed wavelength of $\lambda$4959.

In order to ensure that our photometric EW$_{5007,phot}$ values are on the same scale as spectroscopic EWs (EW$_{5007,spec}$), we compare our derived values of EW$_{5007,phot}$ with reliably measured spectroscopic EWs. 
Motivated by deriving accurate direct chemical abundances, we have obtained high-quality spectra for 28 SFACT GP candidates with the Low Resolution Spectrograph 2 \citep[LRS2;][]{Chonis2016} on the 10~m~Hobby-Eberly Telescope \citep[HET;][]{Ramsey1998,Hill2021}. These galaxies were selected as HET targets solely by their appearance in SFACT Hydra spectra. We note that these targets were selected as GP candidates prior to the development of the GP selection criteria presented in this work. Given the limitations of the Hydra spectra discussed previously, we no longer consider all 28 to be GPs. The analysis of these spectra will be presented in Baker et al. (2026, in preparation). The current work utilizes observed EW$_{5007, spec}$ values measured from the HET spectra solely to calibrate the photometric EW$_{5007,phot}$ values for each filter. In most cases, the continuum is robustly detected, resulting in reliable measurements of spectroscopic EW$_{5007, spec}$ for each galaxy.

Figure \ref{fig:HET_scale} shows the EW$_{5007,spec}$ values measured from the HET spectra versus the EW$_{5007,phot}$ values derived from the BB and NB photometry for each filter. The scale factors for each filter are determined from the slope of a linear least-squares fit to the data, with the intercept fixed at EW$_{5007}$ of zero. These scale factors are broadly consistent with the ratios of the widths of the relevant BB and NB filters. We find scale factors of 13.3~$\pm$~1.3, 13.0~$\pm$~1.0, and 13.1~$\pm$~0.4 for the NB659, NB695, and NB746 filters, respectively. The linear fits are constrained to pass through EW$_{5007}$ of zero because sources with no emission should yield EW$_{5007}$ of zero in both spectroscopic and photometric methods, even if they are not detected by SFACT. As a result, the scale factors are equivalent to the mean ratio of EW$_{5007, spec}$ to EW$_{5007, phot}$. We also show the residuals about the fits in the subplots at the bottom of Figure \ref{fig:HET_scale}. The RMS scatter about the best fit lines are 105~$\text{\AA}$, 81~$\text{\AA}$, and 21~$\text{\AA}$ for the NB659, NB695, and NB746 filters, respectively.

We stress that our derived photometric EW$_{5007,phot}$ values are not meant to be highly precise estimates of the actual EWs in an absolute sense.  For example, our estimates of the continuum fluxes are based on the broadband magnitudes which themselves include the emission lines present in the BB filters. The derived EW$_{5007,phot}$ for the GP candidates presented in this work are tabulated in Section \ref{subsubsec:final_sample}. We calculate the formal uncertainties in EW$_{5007,phot}$ by propagating the photometric uncertainties in the line and BB fluxes, along with the uncertainty in the scaling factors illustrated in Figure \ref{fig:HET_scale}. The EW$_{5007,phot}$ values have uncertainties of \textit{order} 10\%. Nevertheless, they should be reliable relative estimates of the EWs of the [O~III]$\lambda$5007 lines in our objects, which is how they are used in the following section to select GP candidates from the SFACT catalogs.  Furthermore, since they are scaled to reliably measured spectroscopic EWs, our photometric EW$_{5007,phot}$ values should allow for reasonable comparisons with other samples of GP galaxies.

\begin{figure*}[t!]
\centering
\includegraphics[width=\linewidth]{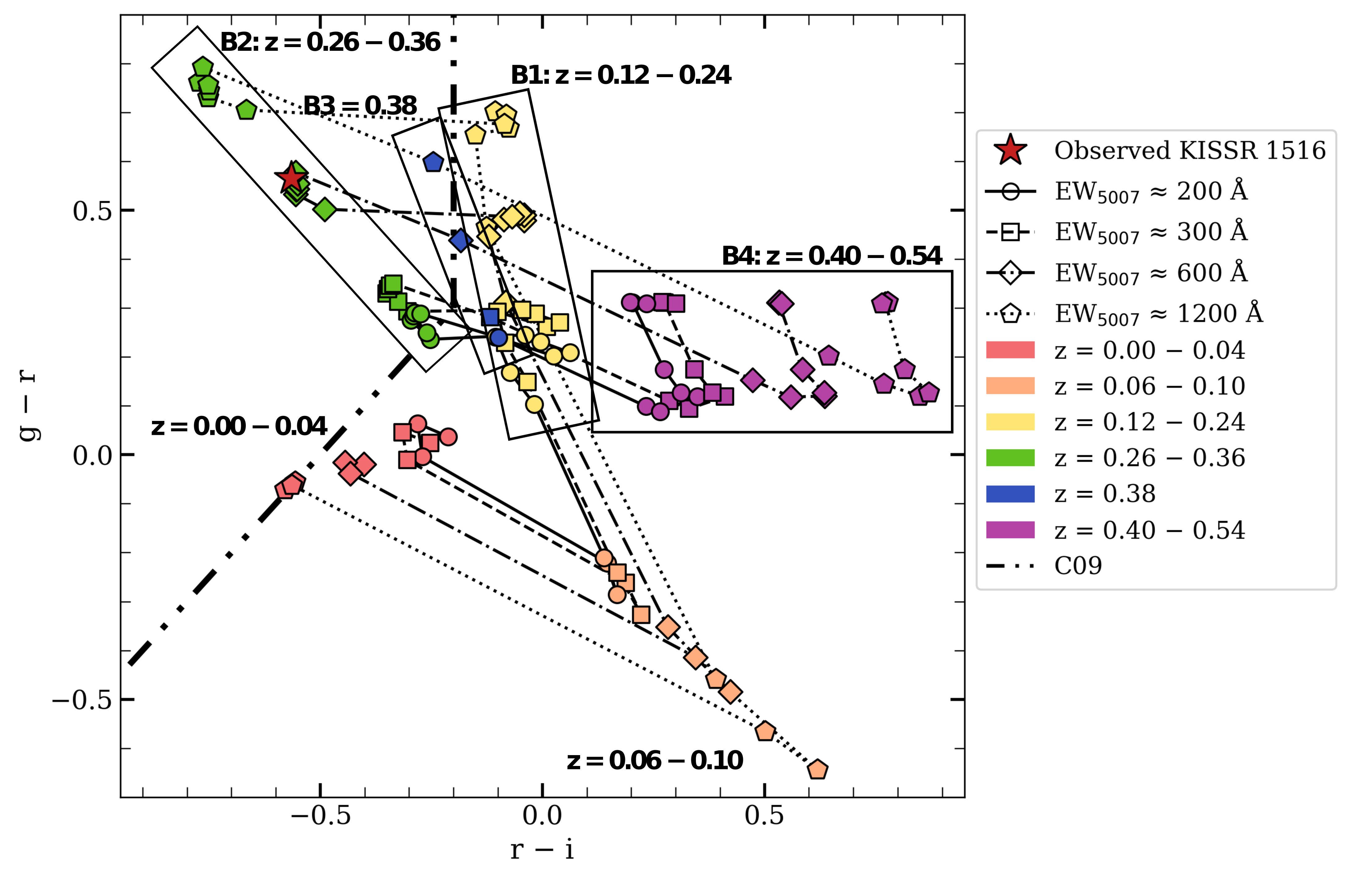} 
\caption{Color-color diagram showing simulated ODI $g - r$ and  $r - i$ colors of four GP template spectra with varying EW$_{5007,spec}$, plotted as redshift tracks over z~=~0~--~0.54 with steps of $\Delta$z~=~0.02. The template GP spectra have EW$_{5007,spec}$~$\approx$~200~$\text{\AA}$ (solid line with circles), 300~$\text{\AA}$ (dashed line with squares), 600~$\text{\AA}$ (dash-dotted line with diamonds), 1200~$\text{\AA}$ (dotted line with pentagons). The markers are color coded by redshift. The template spectra were derived from the observed spectrum of GP KISSR 1516 \citep{Brunker2020}, which has EW$_{5007,spec}$~$\approx$~600~$\text{\AA}$; the corresponding template track passes through the observed $g - r$ and  $r - i$ colors of KISSR 1516 (red star). The boxes highlight redshift ranges at which key spectral transitions occur. B1 highlights z~=~0.12~--~0.24 where [\ion{O}{3}] and H$\alpha$ are in the ODI \textit{r} filter. B2 highlights z~=~0.26~--~0.36 where [\ion{O}{3}] is in the ODI \textit{r} filter, but H$\alpha$ has moved into the \textit{i} filter. B3 highlights z~=~0.38 where [\ion{O}{3}] straddles the ODI \textit{r} and \textit{i} filters. B4 highlights z~=~0.40~--~0.64, where [\ion{O}{3}] moves through the ODI \textit{i} filter. The black dash-dotdotted line represents the \textit{gri} color criteria established in \citet{Cardamone2009} to identify GPs in SDSS at z~$\lesssim$~0.36.}
\label{fig:red_track_odi}
\end{figure*}

\section{Selecting Green Peas in SFACT} \label{sec:SFACT_GPs}
The extreme properties of GPs make them easily identifiable in large surveys (\citetalias{Cardamone2009}; \citealt{Izotov2011}; \citealt{Brunker2020}; \citealt{Liu2022}; \citealt{Ding2023}; \citealt{Kimsey-Miller2024}; \citealt{Jeong2025}; \citealt{Samonski2025}). \citetalias{Cardamone2009} developed robust \textit{ugriz} color selection criteria to catalog the original GP sample in SDSS. The BB color selection method works well to identify \citetalias{Cardamone2009} GPs in the z~=~0.14~--~0.35 window because BB colors, particularly $r - i$, are highly sensitive to the ultra-high EW [\ion{O}{3}] lines present in GP spectra. In this redshift range, the [\ion{O}{3}]$\lambda$5007 line brackets the 50$\%$ throughput range of SDSS~\textit{r} filter, giving a boost to the \textit{r} flux. This boost in the \textit{r} filter leads to the GPs being separated from the main locus of galaxies at the same redshifts when plotted in $g - r$ and $r - i$ space.

The primary goal of the current study is to isolate a sample of GP candidates from among the 1260 [\ion{O}{3}]-detected SFACT galaxies described in Section~\ref{sec:data}.  In this section we lay out our selection methodology, utilizing criteria similar to C09 when possible, and developing new criteria when necessary.  An obvious case of the latter is when the redshifts of the [\ion{O}{3}]-selected SFACT galaxies fall outside the redshift range applicable to the C09 criteria. It is worth stressing that, even though our starting point for creating a sample of SFACT GP candidates is the 1260 galaxies that were detected via their [\ion{O}{3}] lines, only the minority are expected to have emission lines strong enough to qualify them as GPs.  That is, only SFACT sources that exhibit the strongest [\ion{O}{3}] emission should be expected to be selected as GP analogs.

The \textit{gri} criteria developed by \citetalias{Cardamone2009} work well for identifying SDSS GPs in the redshift range where [\ion{O}{3}]$\lambda$5007 falls in the \textit{r} filter. However, at higher redshifts, z~$\gtrsim$~0.36, \textit{gri} color selection criteria for GPs have not been developed, and it is not known \textit{a priori} whether such criteria would be effective in identifying GPs at these higher redshifts. In order to address this question, we investigate the evolution of GP $g - r$ and $r - i$ colors with redshift for a characteristic GP. In Figure \ref{fig:red_track_odi}, we show redshift tracks in \textit{gri} color-color space for four model GP spectra. This allows us to probe the effect on the $g - r$ and $r - i$ colors due to the shifting of the [\ion{O}{3}]$\lambda$5007 line in to and out of the \textit{r} filter.

We simulated the redshift evolution of \textit{gri} colors of GPs by redshifting four GP template spectra from z~=~0 to z~=~0.54 in steps of $\Delta$z~=~0.02 and sampling the spectra through the ODI \textit{gri} filter throughput curves. The template spectra were derived from an observed HET spectrum of the \textit{bona fide} GP KISSR 1516 which has z~=~0.328 and EW$_{5007,spec}$~$\approx$~600~$\text{\AA}$ (\citealt{Brunker2020}; hereafter \citetalias{Brunker2020}). We artificially increased or decreased the strength of the [\ion{O}{3}] $\lambda$5007 line to produce template spectra with observed EW$_{5007,spec}$ of approximately 200, 300, and 1200~$\text{\AA}$. We then scaled the strengths of [\ion{O}{2}], [\ion{O}{3}] $\lambda$4959, H$\beta$, and H$\alpha$ to maintain the expected emission-line ratios observed in the original spectrum. The redshift tracks are presented in Figure \ref{fig:red_track_odi}, where the solid line with circles represents the track of the EW$_{5007,spec}$~$\approx$~200~$\text{\AA}$ template spectrum, the dashed line with squares represents the EW$_{5007,spec}$~$\approx$~300~$\text{\AA}$ template spectrum, and the dotted line represents the EW$_{5007,spec}$~$\approx$~1200~$\text{\AA}$ template spectrum. The original spectrum is represented by the dash-dotted line with diamonds.

\begin{figure*}[t!]
\centering
\includegraphics[width=\linewidth]{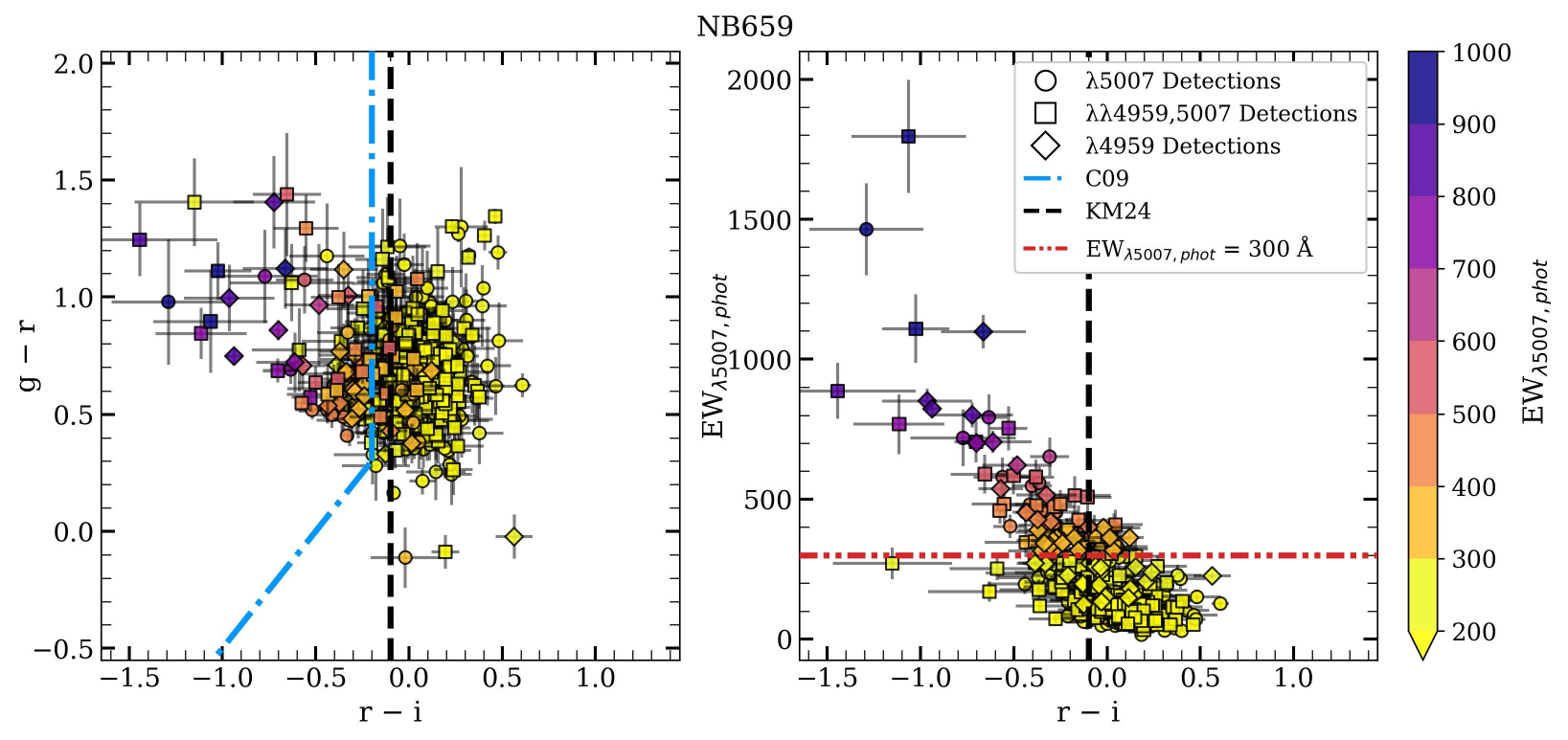} 
\caption{The $g - r$ vs $r - i$ colors (left) and EW$_{5007,phot}$ vs $r - i$ (right) for SFACT NB659 [\ion{O}{3}] detections. In both plots, the data points are color coded by the EW$_{5007,phot}$ value of the galaxy as indicated by the color bar. The blue dash-dotted line in the left panel represents the \textit{gri} color criteria for selecting SDSS GPs established by \citet{Cardamone2009}. The black vertical lines represent the $r - i$ criteria adopted by \citet{Kimsey-Miller2024}. The red horizontal line in the right panel represents the EW$_{5007,phot}$~=~300~$\text{\AA}$ criterion used in this work.}
\label{fig:NB659_gr_vsri_EW_subplots}
\end{figure*}

The data points are color coded by redshift bins, with the bin edges chosen to correspond to the redshifts at which an emission line moves into or out of a broadband filter. We highlight regions of interest with boxes B1 -- B4. The yellow markers in B1 represent the colors produced from the spectra when shifted to z~=~0.12~--~0.24. In this redshift range, the [\ion{O}{3}]$\lambda$$\lambda$4959,5007 and H$\beta$ lines fall in the ODI \textit{r} filter while the H$\alpha$ line falls in the ODI \textit{i} filter. It is worth noting that the simulated colors produced by the template GPs fall outside the color selection criteria established by \citetalias{Cardamone2009} to identify GPs in this redshift range. We note that our results shown in Figure \ref{fig:red_track_odi} may not agree precisely with the color-color trends found using SDSS photometry (e.g., \citetalias{Cardamone2009}) since the SDSS and ODI \textit{gri} filters are not identical. This likely explains why \citetalias{Cardamone2009} could detect GPs at z~=~0.12~--~0.24 using SDSS photometry when they would appear to be excluded based on our analysis utilizing ODI \textit{gri} bandpasses. 

The green markers in B2 represent the colors produced by the template GPs at z~=~0.26~--~0.36. At these redshifts, the [\ion{O}{3}]$\lambda$$\lambda$4959,5007 and H$\beta$ lines fall in the ODI \textit{r} and H$\alpha$ has shifted out of the ODI \textit{i} filter. The red star in B2 represents the observed \textit{gri} colors of  KISSR 1516 as measured from SFACT ODI images. The simulated colors produced by the original spectrum of KISSR 1516, with an EW$_{5007,spec}$ of 600~$\text{\AA}$, overlap with the observed colors of KISSR 1516. We note that the \citetalias{Cardamone2009} color criteria select SDSS GPs in this redshift range, and the GPs simulated colors for z~=~0.26~--~0.36 are within the \citetalias{Cardamone2009} color criteria. B3 highlights a transition region at z~$\approx$~0.38 where the [\ion{O}{3}] lines are redshifting out of the ODI \textit{r} filter and into the ODI \textit{i} filter. This region overlaps in redshift with the NB695 [\ion{O}{3}] detections. 
B4 indicates the redshifts at which the [\ion{O}{3}] lines have fully redshifted into the ODI \textit{i} filter at z~$\gtrsim$~0.4. We discuss B2, B3, and B4 further in the following sections.

While not shown in Figure \ref{fig:red_track_odi}, we applied the same procedure described above to simulate the $g - r$ and $r - i$ colors using SDSS bandpasses. The resulting SDSS redshift tracks have a slight offset towards redder $r - i$  colors, however, they are largely consistent with those presented in \citet{Yang2017...blueberry}. Our SDSS track for the original KISSR 1516 spectrum passes through the observed colors of KISSR 1516 in SDSS. We therefore attribute this offset to differences in how template spectra were constructed. In contrast to the smooth behavior of the SDSS tracks, the ODI $r - i$ tracks exhibit large jumps and sharp inflections in \textit{gri} color space. This is likely due to the ODI filters having sharper edges in the transmission curves, making the broadband colors measured from ODI more sensitive to emission lines moving into or out of the filter bandpasses. Based on the ODI redshift tracks, shown in Figure \ref{fig:red_track_odi}, we find that the \textit{gri} criteria developed by \citetalias{Cardamone2009} should work well to select SFACT NB659 GP candidates at z~=~0.31~--~0.34. However, because the [\ion{O}{3}]$\lambda$5007 line shifts out of the ODI \textit{r} filter for z~$\gtrsim$~0.36, we can not use the \citetalias{Cardamone2009} criteria to identify NB695 or NB746 GP candidates in SFACT.

\subsection{NB659 GP Candidate Sample Selection} \label{subsec:NB659_samp}

Galaxies detected by $\lambda$4959, $\lambda$5007, or both via the NB659 filter fall in the redshift range, z~=~0.31~--~0.34, thus their [\ion{O}{3}] lines fall within the ODI \textit{r} filter. This redshift window of NB659 [\ion{O}{3}] detections falls within B2 of Figure \ref{fig:red_track_odi}. Further, SFACT GP candidates at these redshifts fall within the redshift range of GPs discovered by \citetalias{Cardamone2009} and are expected to satisfy the \citetalias{Cardamone2009} color criteria. In Figure \ref{fig:NB659_gr_vsri_EW_subplots}, we show the $g - r$ vs $r - i$ colors with their errors for the full sample of NB659 [\ion{O}{3}]-detected SFACT galaxies.  Here and in all subsequent plots that present SFACT colors and magnitudes, the photometry has been corrected for Galactic extinction using absorption and reddening corrections from \citet{schlafly2011}.  Each data point is color coded by the EW$_{5007,phot}$ measured for the galaxy, with darker colors representing stronger [\ion{O}{3}] emission. The blue dash-dotted lines represent the \citetalias{Cardamone2009} $g - r$ and $r - i$ color criteria. The black dashed line represents the $r - i$ color criterion modified from the \citetalias{Cardamone2009} $r - i$ criterion adopted by \citet[][hereafter \citetalias{Kimsey-Miller2024}]{Kimsey-Miller2024}. \citetalias{Kimsey-Miller2024} extended the color criterion to include [\ion{O}{3}]-detected H$\alpha$ Dots \citep{Kellar2012,Salzer2020,Watkins2021} with GP-like spectra that were excluded by the \citetalias{Cardamone2009} criteria. We adopt the $r - i \leq -0.1$ criterion established by \citetalias{Kimsey-Miller2024} as our color criterion for selecting SFACT GP candidates in the NB659 filter.

We also plot the EW$_{5007,phot}$ vs $r - i$ color for the NB659 [\ion{O}{3}] detections in Figure \ref{fig:NB659_gr_vsri_EW_subplots}. The black line represents the \citetalias{Kimsey-Miller2024} $r - i$ criterion adopted in this work. In this redshift window, there is a strong correlation between $r - i$ color and EW$_{5007,phot}$, with high EW$_{5007,phot}$ sources trending towards bluer $r - i$ colors. While \citetalias{Cardamone2009} GPs appear to be offset from the main locus of galaxies in \textit{gri} color-color space, the SFACT NB659 [\ion{O}{3}]-detected sources form a continuum in $r - i$. For this reason, we apply a EW$_{5007,phot}$ limit of 300~$\text{\AA}$ to exclude weak-lined objects from our sample. This is represented by the red horizontal line in the right-hand panel of Figure \ref{fig:NB659_gr_vsri_EW_subplots}. 

We note that our adopted EW$_{5007,phot}$ limit of 300~$\text{\AA}$ is fairly conservative since there exist galaxies exhibiting characteristic GP properties identified with EW$_{5007,spec}$ values that extend below this threshold (\citetalias{Cardamone2009};  \citetalias{Brunker2020}; \citealt{Jeong2025}). Our EW$_{5007,phot}$ limit and the \citetalias{Kimsey-Miller2024} color criterion result in 65 NB659 GP candidates. 

\begin{figure*}[t!]
\centering
\includegraphics[width=\linewidth]{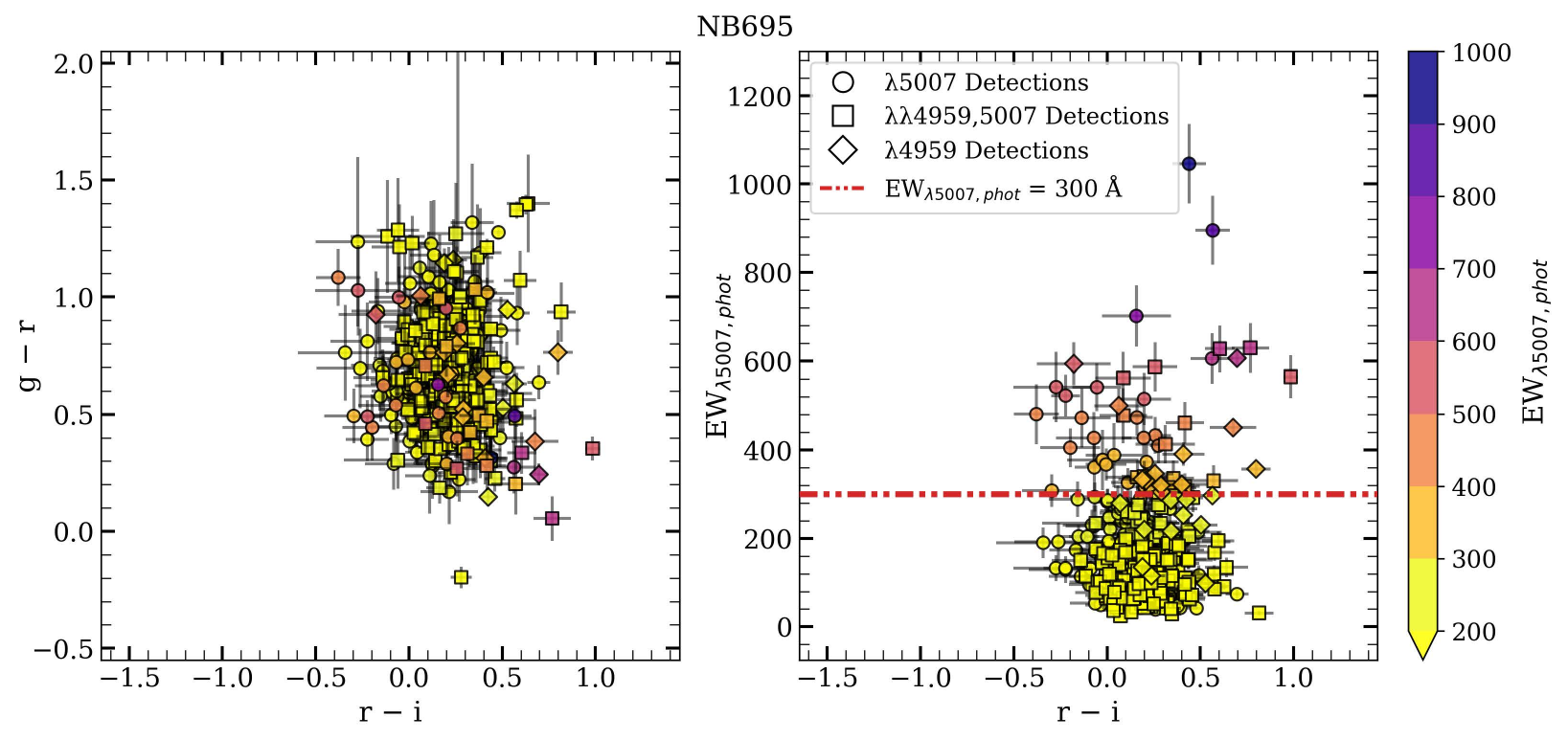} 
\caption{The $g - r$ vs $r - i$ colors (left) and EW$_{5007,phot}$ vs$r - i$ (right) for NB695 [\ion{O}{3}]-detected SFACT galaxies. The data points are color coded by the EW$_{5007,phot}$ value of the galaxy, indicated with the color bar. The EW$_{5007,phot}$~=~300~$\text{\AA}$ criterion adopted in this work is represented by the red horizontal line.}
\label{fig:NB695_gr_vsri_EW_subplots}
\end{figure*}

\subsection{NB695 GP Candidate Sample Selection} \label{subsec:NB695_samp}

The NB695 [\ion{O}{3}] detections fall at z~=~0.38~--~0.41. We note that GP-like galaxies at z~$\gtrsim$~0.36 are sometimes referred to as ``Purple Grapes" (\citetalias{Brunker2020}; \citealt{Liu2022}). \citetalias{Brunker2020} demonstrated that Purple Grapes exhibit the same physical characteristics and emission-line properties as lower-redshift GPs and thus belong to the same galaxy population. In this work, we will still refer to them as GPs. 

As shown in Figure \ref{fig:trans_curves}, the NB695 filter is placed at the edges of the ODI \textit{r} and \textit{i} filters which leads to a wide span in possible $r - i$ color space for SFACT NB695 detections. This is illustrated by the tracks in Figure \ref{fig:red_track_odi}. NB695 GP candidates at the low redshift end of z~$\approx$~0.38 will fall near the B3 region and move towards increasing $r - i$ color with increasing redshift until they reach the B4 region at the higher redshift end with z~$\approx$~0.41. This eliminates the possibility of utilizing \textit{gri} color criteria to isolate the population of GPs detected by NB695 because the GPs do not distinguish themselves adequately from the other star-forming SFACT galaxies in this redshift range. To further illustrate this, we plot the $g - r$ vs $r - i$ colors (left) and EW$_{5007,phot}$ vs $r - i$ color (right) of the NB695 [\ion{O}{3}]-detected sample in Figure \ref{fig:NB695_gr_vsri_EW_subplots} with each point color coded by the galaxy's EW$_{5007,phot}$. Because the NB695 filter covers a wavelength range corresponding to the edges of the ODI \textit{r} and \textit{i} filter, the $r - i$ color of the GPs depends highly on its redshift and, thus, where the [\ion{O}{3}] line falls in either BB filter. Where the SFACT galaxies show a clear correlation between the $r - i$ color and EW$_{5007,phot}$ in Figure \ref{fig:NB659_gr_vsri_EW_subplots}, the galaxies show no such trend in Figure \ref{fig:NB695_gr_vsri_EW_subplots}: the EW$_{5007,phot}$ values are essentially independent of $r - i$ color. 

\begin{figure*}[t!]
\centering
\includegraphics[width=\linewidth]{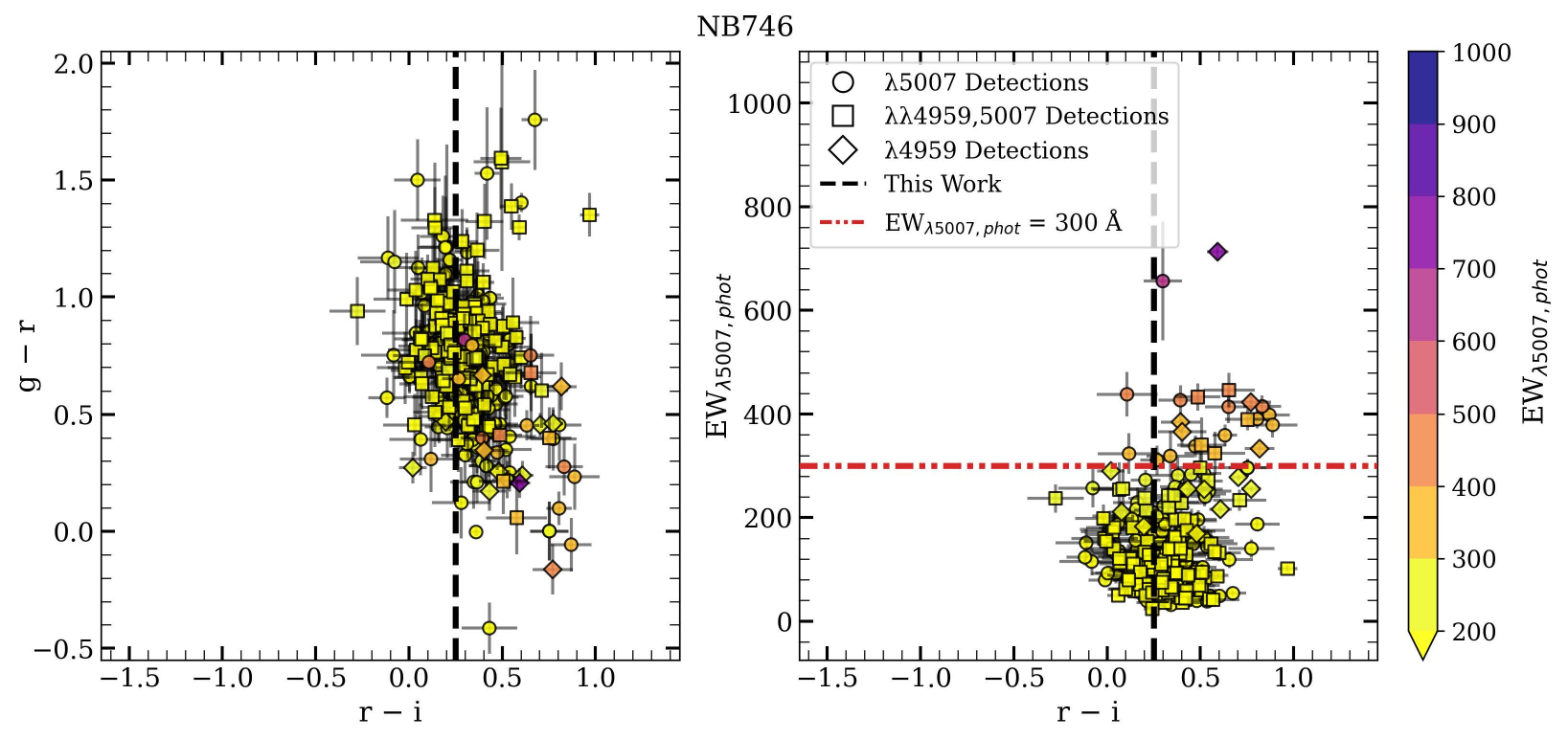} 
\caption{The $g - r$ vs $r - i$ colors (left) and EW$_{5007,phot}$ vs $r - i$ (right) for NB746 [\ion{O}{3}] detections. The color of the points indicated on the color bar represent the EW$_{5007,phot}$ values of the galaxies. The black vertical lines represent the $r - i$ criteria adopted in this work to select GPs in z~=~0.48~--~0.51. The red horizontal line represents the EW$_{5007,phot}$~=~300~$\text{\AA}$ criterion.}
\label{fig:NB746_gr_vsri_EW_subplots}
\end{figure*}

Hence, we cannot use a color criterion to distinguish strong [\ion{O}{3}] emitters for this filter. Instead we are limited to using EW$_{5007,phot}$ as our sole NB695 selection criterion. We adopt the EW$_{5007,phot}$ limit of 300~$\text{\AA}$, shown as the red line in the righthand plot in Figure \ref{fig:NB695_gr_vsri_EW_subplots}. This EW$_{5007,phot}$ limit was chosen to mirror the limit used for the NB659 GP candidates. With this EW$_{5007,phot}$ limit, we find 51 NB695 GP candidates.

\subsection{NB746 GP Candidate Sample Selection} \label{subsec:NB746_samp}

The NB746 sample of [\ion{O}{3}] detections falls in the z~=~0.48~--~0.51 range. At this redshift range, the [\ion{O}{3}] line has shifted into the ODI \textit{i} filter. The extreme [\ion{O}{3}] emission will give a boost to the flux in \textit{i} and push the $r - i$ color of GPs to larger values, as seen in B4 in Figure \ref{fig:red_track_odi}. Color selection criteria do not exist for GPs at these redshifts, so we develop a $r - i$ color criterion for SFACT GP candidates at z~=~0.48~--~0.51. We again maintain the EW$_{5007,phot}$ limit of 300~$\text{\AA}$ used for NB659 and NB695 to exclude weak-lined objects. 

We obtain the NB746 color selection criterion by redshifting a template GP spectrum with EW$_{5007,spec}$ of 300~$\text{\AA}$ to z~=~0.48~--~0.51 and simulating the expected $r - i$ colors at these redshifts, as shown in Figure \ref{fig:red_track_odi}. We find GPs with EW$_{5007}$~$\geq$~300~$\text{\AA}$ at these redshifts have $r - i$~$\geq$~0.25. We adopt this as our color criterion for the NB746 GP sample. In Figure \ref{fig:NB746_gr_vsri_EW_subplots}, we plot the $g - r$ vs $r - i$ colors (left) and EW$_{5007,phot}$ vs $r - i$ color (right) of the NB746 [\ion{O}{3}] detections. Each point is color coded based on how strong the EW$_{5007,phot}$ is, with darker points representing larger EW$_{5007,phot}$ values. The $r - i$ color criterion for this filter is represented as a black line. The right hand panel of Figure \ref{fig:NB746_gr_vsri_EW_subplots} shows a weak correlation between $r - i$ and EW$_{5007,phot}$, with the majority of NB746 galaxies with high EW$_{5007,phot}$ trending towards towards larger $r - i$  colors. We find that 20 NB746 GP candidates satisfy our $r - i$ criterion and EW$_{5007,phot}$ limit. 

\subsection{Reducing Contaminants} \label{subsec:contam}
Other classes of extragalactic objects, mainly Seyfert 2 (Sy2) galaxies, can exhibit strong, narrow, [\ion{O}{3}] emission and hence could represent contaminants within our sample. During the SFACT spectral processing, objects with broad emission lines like Seyfert 1 (Sy1) galaxies and quasars are flagged and hence not considered as possible GP candidates. Thus, there is no contamination of broad line systems in our sample. We also do not expect to have any contamination from narrow-line Sy1 (NLS1) galaxies in our sample. \citetalias{Cardamone2009} classified objects with $500\leq \text{FWHM}_{H \beta} \leq 2000$~km~s$^{-1}$ in their sample as NLS1. The SFACT GP candidates presented in this work have $\text{FWHM}_{H \beta} < 300$~km~s$^{-1}$, and thus, are not expected to be NLS1 objects. 

\begin{figure*}[t!]
\centering
\includegraphics[width=0.9\linewidth]{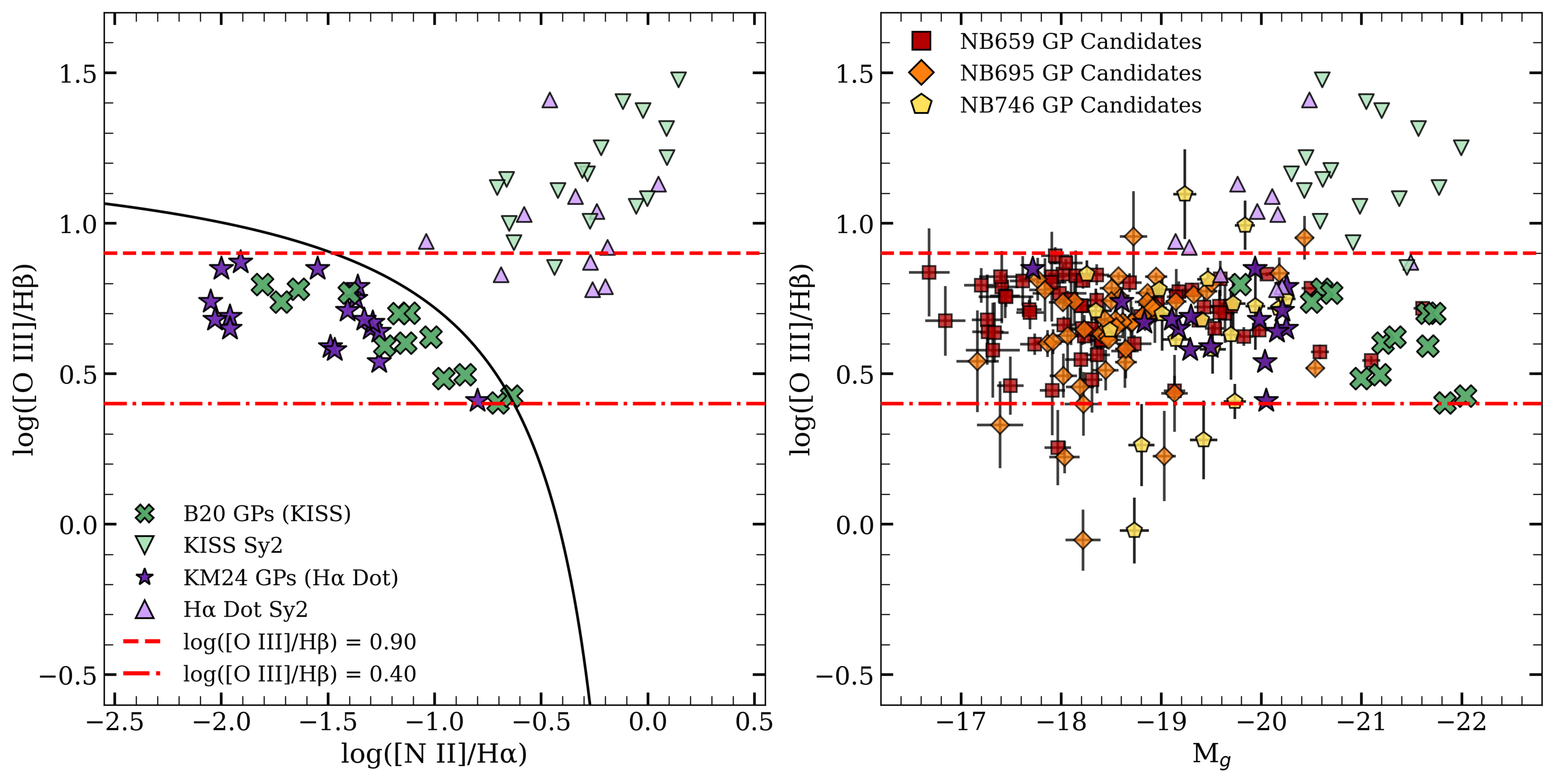} 
\caption{\textit{Left:} Emission-line diagnostic diagram utilizing [\ion{O}{3}]/H$\beta$ and [\ion{N}{2}]/H$\alpha$ of GPs and Seyfert 2 galaxies from the KISS and H$\alpha$ Dot Surveys. The solid line represents the \citet{Kauffmann2003} relation used to distinguish between AGN and SFGs. The Green Peas fall between log([\ion{O}{3}]/H$\beta$) values of 0.4 and 0.9, represented by red dash-dotted and dashed lines, respectively. Most, but not all, of the Seyfert 2 galaxies have log([\ion{O}{3}]/H$\beta$)~$>$~0.9. \textit{Right:} Log([\ion{O}{3}]/H$\beta$) vs g-band absolute magnitude values of the same KISS and H$\alpha$ Dot GPs and Sy2s. The SFACT GP samples are also plotted. SFACT objects with log([\ion{O}{3}]/H$\beta$)~$<$~0.4 or log([\ion{O}{3}]/H$\beta$)~$>$~0.9 are excluded from the SFACT GP samples.}
\label{fig:bpt_mg_plot_subplot}
\end{figure*}


The ratios of strong emission lines serve as key diagnostic tools for identifying the primary source of ionizing photons present in ELGs with narrow emission lines \citep{Baldwin1981,Veilleux1987}. Most commonly, the Baldwin, Phillips, and Terlevich diagram \citep[BPT;][]{Baldwin1981}, which utilizes [\ion{O}{3}]/H$\beta$ and [\ion{N}{2}]/H$\alpha$, is used to distinguish between SFGs and AGNs, mainly Sy2 galaxies, in ELG samples. While SFACT employs this diagnostic for the low-redshift H$\alpha$ detections, the H$\alpha$ and [\ion{N}{2}] lines are redshifted beyond the range of the spectral coverage for all SFACT [\ion{O}{3}] detections. 

Diagnostic diagrams utilizing the blue emission lines (i.e., [\ion{O}{2}]~$\lambda$3727, [\ion{O}{3}]~$\lambda$5007, and H$\beta$) have been proposed in the literature (\citealt{Baldwin1981}; \citealt{Lamareille2004}). However, because of the long wavelength baseline, the ratios between [\ion{O}{2}]~$\lambda$3727 and [\ion{O}{3}]~$\lambda$5007 or H$\beta$ are subject to reddening affects (\citealt{Perez-montero2013}). We note that corrections have been proposed (\citealt{Perez-montero2013}, references therein), however, the limitations of our spectra (See Section \ref{sec:O3_det}) eliminate this as an option. Hence, an alternate method must be applied to identify and remove Sy2s from our sample of GP candidates. 

In the left panel of Figure \ref{fig:bpt_mg_plot_subplot}, we show a BPT diagram of GPs (\citetalias{Brunker2020}; \citetalias{Kimsey-Miller2024}) and Sy2 galaxies from the KPNO International Spectroscopic Survey (KISS; \citealt{Salzer2000}, \citealt{Salzer2001}, \citealt{Wegner2003}, \citealt{Gronwall2004a}, \citealt{Jangren2005}, \citealt{Salzer2005}) and H$\alpha$ Dot Survey (\citealt{Kellar2012}, \citealt{Salzer2020}). In the right panel of Figure \ref{fig:bpt_mg_plot_subplot}, we plot log([\ion{O}{3}]/H$\beta$) against the g-band absolute magnitudes for the same KISS and H$\alpha$ Dot GP and Sy2 samples. The absolute magnitudes for the KISS and H$\alpha$ Dot galaxies were calculated using SDSS photometry. We also show the SFACT NB659, NB695, and NB746 GP samples in the right panel of Figure \ref{fig:bpt_mg_plot_subplot}. We note that the absolute magnitudes used in the current work have not been k-corrected. However, we expect the redshift-dependent effects to be small for g-band absolute magnitude in the redshift windows probed in this study.

All KISS and H$\alpha$ Dot GPs have log([\ion{O}{3}]/H$\beta$) less than 0.9, which is shown as red dashed line in both plots. While there is a modest overlap in log([\ion{O}{3}]/H$\beta$) for the KISS and H$\alpha$ Dot GPs and Sy2 galaxies, most Sy2s have log([\ion{O}{3}]/H$\beta$) greater than 0.9. Therefore, we exclude SFACT GP candidates with log([\ion{O}{3}]/H$\beta$)~$\geq$~0.9, including two NB695 and two NB746 GP candidates, as possible Sy2 contaminants. 

\begin{table}
\caption{SFACT GP Selection Criteria}
\footnotesize
\begin{center}
    \begin{tabular}{ccccc} 
    \hline\hline
    SFACT &  $r - i$ & EW$_{5007,phot}$ & log($\frac{[O III]}{H\beta}$) & No. GP \\
    Filter &   &   &   & Candidates \\
    & & ($\text{\AA}$) & & \\
    \hline
    NB659 & $\leq$ $-0.10$ & $\geq$ 300 & 0.4 -- 0.9 & 64\\
    NB695 & --- & $\geq$ 300 & 0.4 -- 0.9 & 44\\
    NB746 & $\geq$ 0.25 & $\geq$ 300 & 0.4 -- 0.9 & 15\\
    \hline 
    \end{tabular}
\end{center}
\label{tab:criteria}
\end{table}

\noindent The KISS and H$\alpha$ Dot GPs and Sy2s that overlap in log([\ion{O}{3}]/H$\beta$) also overlap in $M_{g}$. Most of the SFACT GP candidates in this overlapping range of log([\ion{O}{3}]/H$\beta$), are less luminous than the Sy2s. We stress that SFACT samples lower luminosity sources.


\noindent than possible with the KISS and H$\alpha$ Dot surveys. For this reason, we do not use these samples to apply an absolute magnitude cut in favor of keeping lower-luminosity sources in our sample. We discuss this further in Section \ref{subsec:low_luminosity}. 

The KISS and H$\alpha$ Dot GPs also have log([\ion{O}{3}]/H$\beta$) greater than 0.4, shown as a red dot-dashed line in both plots, indicating that they are high-excitation systems. We apply a limit of log([\ion{O}{3}]/H$\beta$)~$\geq$~0.4 to exclude low-excitation objects from our sample that are unlikely to be GPs. We exclude one NB659, five NB695, and three NB746 GP candidates that fall below this line. While our selection criteria outlined above are effective at removing most Sy2 contaminants, it is still possible that a small number of Sy2 AGNs have been included in our GP catalog.  Hence, we will continue to refer to them as GP candidates. We note that we are in the process of obtaining additional spectra of our GP candidates that include the redshifted [\ion{N}{2}] and H$\alpha$ lines. In future studies of this sample, we expect to be able to utilize BPT diagnostics to properly distinguish between star-forming GP candidates and Sy2 contaminants.

\subsection{Final Sample} 

\label{subsubsec:final_sample}
We list the selection criteria detailed in the sections above for each NB filters in Table \ref{tab:criteria}. Our final GP sample consists of 64 NB659 GP candidates, 44 NB695 GP candidates, and 15 NB746 GP candidates. The smaller sample numbers reported here compared to Sections \ref{subsec:NB659_samp}, \ref{subsec:NB695_samp}, and \ref{subsec:NB746_samp} reflect objects lost due to our log([\ion{O}{3}]/H$\beta$) criteria. We investigate the properties

\begin{longrotatetable}
\begin{deluxetable}{lccccrrcccc}
\tablecaption{NB659 SFACT Green Peas Candidates}
\tablewidth{0pt}
\tabletypesize{\scriptsize}
\tablehead{
\colhead{SFACT Object ID} & \colhead{$\alpha$(J2000)} & \colhead{$\delta$(J2000)} & \colhead{\textit{z}} &
\colhead{$m_{r}$} & \colhead{$r - i$} & \colhead{$g - r$} &
\colhead{EW$_{5007,phot}$} & \colhead{$M_{g}$}& \colhead{log([O III]/H$\beta$)} & \colhead{log(SFR)}
\\
\colhead{ } & \colhead{(deg)} & \colhead{(deg)} & \colhead{} &
\colhead{(mag)} & \colhead{(mag)} & \colhead{(mag)} &
\colhead{($\text{\AA}$)} & \colhead{(mag)}&\colhead{} & \colhead{(M$_{\odot}$ yr$^{-1}$)}
\\
\colhead{(1)} & \colhead{(2)} & \colhead{(3)} & \colhead{(4)} &
\colhead{(5)} & \colhead{(6)} & \colhead{(7)} &
\colhead{(8)} & \colhead{(9)}&\colhead{(10)}& \colhead{(11)}
} 
\startdata
SFF07$-$NB2$-$A7587$^{b}$ & 2.93384 & 26.40837 & 0.33304 &$21.38 \pm 0.04$ & $-0.18 \pm 0.08$ & $0.68 \pm 0.07$ & $362 \pm 16$ & $-19.18 \pm 0.10$ & $0.772 \pm 0.034$ & $1.309 \pm 0.047$\\
SFF08$-$NB2$-$B10370 & 10.42952 & 25.63743 & 0.31970 &$22.74 \pm 0.07$ & $-0.10 \pm 0.13$ & $0.78 \pm 0.12$ & $507 \pm 62$ & $-17.61 \pm 0.13$ & $0.809 \pm 0.082$ & $0.694 \pm 0.044$\\
SFF08$-$NB2$-$A9786 & 10.77276 & 25.58735 & 0.31119 &$22.00 \pm 0.07$ & $-0.15 \pm 0.12$ & $0.71 \pm 0.12$ & $303 \pm 38$ & $-18.36 \pm 0.13$ & $0.562 \pm 0.128$ & $0.813 \pm 0.046$\\
SFF09$-$NB2$-$B9603 & 20.66606 & 27.83047 & 0.32648 &$22.17 \pm 0.05$ & $-0.19 \pm 0.10$ & $0.67 \pm 0.08$ & $389 \pm 45$ & $-18.35 \pm 0.11$ & $0.744 \pm 0.027$ & $0.855 \pm 0.042$\\
SFF09$-$NB2$-$C25530 & 21.05726 & 28.16829 & 0.33082 &$21.14 \pm 0.03$ & $-0.37 \pm 0.07$ & $0.55 \pm 0.05$ & $427 \pm 16$ & $-19.53 \pm 0.10$ & $0.651 \pm 0.023$ & $1.419 \pm 0.040$\\
SFF11$-$NB2$-$B12344 & 29.79182 & 24.57154 & 0.32252 &$20.88 \pm 0.04$ & $-0.70 \pm 0.08$ & $0.69 \pm 0.05$ & $705 \pm 74$ & $-19.59 \pm 0.09$ & $0.816 \pm 0.059$ & $1.667 \pm 0.038$\\
SFF12$-$NB2$-$B2392 & 30.02152 & 23.98760 & 0.31542 &$22.67 \pm 0.07$ & $-1.29 \pm 0.30$ & $0.98 \pm 0.27$ & $1464 \pm 165$ & $-17.45 \pm 0.27$ & $0.756 \pm 0.034$ & $0.997 \pm 0.040$\\
SFF12$-$NB2$-$B1771 & 30.04464 & 23.96419 & 0.31857 &$21.72 \pm 0.06$ & $-0.35 \pm 0.12$ & $0.68 \pm 0.16$ & $351 \pm 40$ & $-18.73 \pm 0.17$ & $0.599 \pm 0.028$ & $0.997 \pm 0.041$\\
SFF11$-$NB2$-$C12293$^{b}$ & 30.12376 & 24.90712 & 0.32950 &$19.97 \pm 0.02$ & $-0.94 \pm 0.05$ & $0.75 \pm 0.03$ & $823 \pm 30$ & $-20.49 \pm 0.09$ & $0.784 \pm 0.004$ & $2.176 \pm 0.038$\\
SFF11$-$NB2$-$C5398 & 30.34079 & 25.09438 & 0.32420 &$22.90 \pm 0.09$ & $-0.22 \pm 0.18$ & $1.00 \pm 0.16$ & $380 \pm 54$ & $-17.27 \pm 0.16$ & $0.639 \pm 0.061$ & $0.430 \pm 0.048$\\
SFF13$-$NB2$-$B4867 & 37.39646 & 26.16787 & 0.32296 &$22.27 \pm 0.07$ & $-0.38 \pm 0.15$ & $1.00 \pm 0.14$ & $478 \pm 57$ & $-17.90 \pm 0.15$ & $0.804 \pm 0.033$ & $0.837 \pm 0.043$\\
SFF13$-$NB2$-$A6174 & 37.66320 & 26.13801 & 0.32453 &$22.42 \pm 0.07$ & $-0.66 \pm 0.18$ & $1.44 \pm 0.26$ & $589 \pm 70$ & $-17.32 \pm 0.27$ & $0.578 \pm 0.158$ & $0.807 \pm 0.041$\\
SFF13$-$NB2$-$A1415 & 37.83150 & 26.47477 & 0.30976 &$20.77 \pm 0.03$ & $-0.36 \pm 0.06$ & $0.59 \pm 0.05$ & $558 \pm 58$ & $-19.69 \pm 0.09$ & $0.725 \pm 0.017$ & $1.626 \pm 0.037$\\
SFF14$-$NB2$-$D5817 & 38.46156 & 27.62365 & 0.32421 &$22.67 \pm 0.09$ & $-0.43 \pm 0.22$ & $0.59 \pm 0.13$ & $346 \pm 54$ & $-17.91 \pm 0.13$ & $0.445 \pm 0.149$ & $0.477 \pm 0.053$\\
SFF14$-$NB2$-$A10613$^{b}$ & 38.64331 & 27.28166 & 0.32136 &$20.88 \pm 0.03$ & $-0.25 \pm 0.06$ & $0.68 \pm 0.05$ & $483 \pm 50$ & $-19.58 \pm 0.09$ & $0.703 \pm 0.074$ & $1.576 \pm 0.037$\\
SFF15$-$NB2$-$D22777 & 39.38748 & 28.16978 & 0.31010 &$20.71 \pm 0.04$ & $-0.44 \pm 0.07$ & $0.52 \pm 0.05$ & $453 \pm 47$ & $-19.82 \pm 0.09$ & $0.624 \pm 0.031$ & $1.529 \pm 0.038$\\
SFF15$-$NB2$-$D19839 & 39.42469 & 28.17639 & 0.33183 &$22.39 \pm 0.10$ & $-0.23 \pm 0.17$ & $0.63 \pm 0.14$ & $350 \pm 25$ & $-18.21 \pm 0.14$ & $0.650 \pm 0.055$ & $0.645 \pm 0.074$\\
SFF15$-$NB2$-$B8478 & 39.56128 & 27.60335 & 0.33014 &$21.60 \pm 0.06$ & $-0.30 \pm 0.10$ & $0.48 \pm 0.09$ & $333 \pm 20$ & $-19.13 \pm 0.11$ & $0.443 \pm 0.050$ & $1.000 \pm 0.067$\\
SFF15$-$NB2$-$A6234 & 39.89721 & 27.74993 & 0.32837 &$22.20 \pm 0.06$ & $-0.37 \pm 0.11$ & $0.77 \pm 0.10$ & $389 \pm 21$ & $-18.23 \pm 0.11$ & $0.625 \pm 0.022$ & $0.687 \pm 0.058$\\
SFF15$-$NB2$-$A3355 & 40.00197 & 27.77954 & 0.31616 &$22.59 \pm 0.08$ & $-0.56 \pm 0.15$ & $1.07 \pm 0.14$ & $579 \pm 69$ & $-17.44 \pm 0.15$ & $0.758 \pm 0.073$ & $0.636 \pm 0.042$\\
SFF15$-$NB2$-$C6598 & 40.05092 & 27.86637 & 0.31535 &$21.61 \pm 0.07$ & $-0.40 \pm 0.11$ & $0.62 \pm 0.09$ & $548 \pm 60$ & $-18.87 \pm 0.10$ & $0.729 \pm 0.015$ & $1.154 \pm 0.040$\\
SFF23$-$NB2$-$A1757 & 45.52703 & 19.89396 & 0.32494 &$22.12 \pm 0.06$ & $-1.02 \pm 0.18$ & $1.11 \pm 0.12$ & $1108 \pm 122$ & $-17.94 \pm 0.14$ & $0.892 \pm 0.029$ & $1.056 \pm 0.039$\\
SFS01$-$NB2$-$A4478 & 130.26027 & 13.48780 & 0.33355 &$22.22 \pm 0.07$ & $-0.96 \pm 0.24$ & $0.99 \pm 0.14$ & $851 \pm 44$ & $-18.02 \pm 0.15$ & $0.831 \pm 0.032$ & $1.108 \pm 0.054$\\
SFS02$-$NB2$-$D15636 & 131.55846 & 13.68332 & 0.31823 &$21.19 \pm 0.04$ & $-0.41 \pm 0.08$ & $0.50 \pm 0.06$ & $481 \pm 50$ & $-19.43 \pm 0.10$ & $0.721 \pm 0.034$ & $1.384 \pm 0.038$\\
SFS02$-$NB2$-$C27345$^{b}$ & 131.91994 & 13.45759 & 0.32312 &$21.86 \pm 0.07$ & $-1.44 \pm 0.42$ & $1.24 \pm 0.16$ & $886 \pm 99$ & $-18.05 \pm 0.17$ & $0.868 \pm 0.013$ & $1.205 \pm 0.038$\\
SFS02$-$NB2$-$C19263 & 132.05148 & 13.46468 & 0.32371 &$22.99 \pm 0.10$ & $-0.17 \pm 0.19$ & $0.96 \pm 0.18$ & $513 \pm 70$ & $-17.20 \pm 0.17$ & $0.793 \pm 0.058$ & $0.547 \pm 0.046$\\
SFS02$-$NB2$-$A7068$^{c}$ & 132.09827 & 13.13380 & 0.31138 &$21.06 \pm 0.04$ & $-0.63 \pm 0.10$ & $0.69 \pm 0.06$ & $793 \pm 81$ & $-19.31 \pm 0.10$ & $0.778 \pm 0.023$ & $1.618 \pm 0.037$\\
SFS03$-$NB2$-$C36676$^{c}$ & 132.37646 & 20.24297 & 0.31196 &$20.91 \pm 0.03$ & $-0.52 \pm 0.09$ & $0.52 \pm 0.04$ & $403 \pm 42$ & $-19.64 \pm 0.09$ & $0.700 \pm 0.098$ & $1.461 \pm 0.038$\\
SFS04$-$NB2$-$D5480 & 134.68967 & 5.22653 & 0.32173 &$22.85 \pm 0.09$ & $-1.06 \pm 0.31$ & $0.90 \pm 0.22$ & $1796 \pm 203$ & $-17.40 \pm 0.22$ & $0.822 \pm 0.015$ & $1.039 \pm 0.039$\\
SFS04$-$NB2$-$C14595$^{b}$ & 135.05603 & 4.98361 & 0.33653 &$20.34 \pm 0.03$ & $-0.70 \pm 0.05$ & $0.86 \pm 0.04$ & $701 \pm 26$ & $-20.05 \pm 0.09$ & $0.830 \pm 0.006$ & $2.051 \pm 0.039$\\
\enddata
\tablecomments{This table is available in its entirety in machine-readable form. $a$. KISS GP (\citealt{Brunker2020}) $b$. H$\alpha$Dot GP (\citealt{Kimsey-Miller2024}) \added{$c$. GP candidate (\citealt{Samonski2025})}}
\label{tab: NB659}
\end{deluxetable}
\end{longrotatetable}
\begin{longrotatetable}
\begin{deluxetable}{lccccrrcccc}
\tablecaption{NB695 SFACT Green Peas Candidates}
\tablewidth{0pt}
\tabletypesize{\scriptsize}
\tablehead{
\colhead{SFACT Object ID} & \colhead{$\alpha$(J2000)} & \colhead{$\delta$(J2000)} & \colhead{\textit{z}} &
\colhead{$m_{r}$} & \colhead{$r - i$} & \colhead{$g - r$} &
\colhead{EW$_{5007,phot}$} & \colhead{$M_{g}$}& \colhead{log([O III]/H$\beta$)} & \colhead{log(SFR)}
\\
\colhead{ } & \colhead{(deg)} & \colhead{(deg)} & \colhead{} &
\colhead{(mag)} & \colhead{(mag)} & \colhead{(mag)} &
\colhead{($\text{\AA}$)} & \colhead{(mag)}&\colhead{} & \colhead{(M$_{\odot}$ yr$^{-1}$)}
\\
\colhead{(1)} & \colhead{(2)} & \colhead{(3)} & \colhead{(4)} &
\colhead{(5)} & \colhead{(6)} & \colhead{(7)} &
\colhead{(8)} & \colhead{(9)}&\colhead{(10)}& \colhead{(11)}
} 
\startdata
SFF07$-$NB1$-$B5596 & 2.66143 & 26.37534 & 0.38966 & $ 22.60 \pm0.06 $ & $ 0.26 \pm  0.09 $ & $ 0.40 \pm  0.08 $ & $ 433 \pm  41 $ & $ -18.64 \pm  0.11 $ & $ 0.583 \pm  0.026 $ & $ 1.049 \pm  0.042 $ \\
SFF07$-$NB1$-$B5141 & 2.67489 & 26.20170 & 0.38864 & $ 22.51 \pm0.07 $ & $ 0.20 \pm  0.12 $ & $ 0.57 \pm  0.11 $ & $ 427 \pm  42 $ & $ -18.55 \pm  0.12 $ & $ 0.655 \pm  0.056 $ & $ 1.078 \pm  0.042 $ \\
SFF09$-$NB1$-$D2878 & 20.81356 & 28.17671 & 0.38462 & $ 22.96 \pm0.08 $ & $ -0.14 \pm  0.15 $ & $ 0.62 \pm  0.13 $ & $ 472 \pm  67 $ & $ -18.02 \pm  0.14 $ & $ 0.493 \pm  0.073 $ & $ 0.775 \pm  0.058 $ \\
SFF09$-$NB1$-$C21032 & 21.11762 & 28.28524 & 0.39471 & $ 22.81 \pm0.07 $ & $ 0.35 \pm  0.11 $ & $ 1.03 \pm  0.15 $ & $ 336 \pm  41 $ & $ -17.84 \pm  0.16 $ & $ 0.778 \pm  0.105 $ & $ 0.850 \pm  0.051 $ \\
SFF21$-$NB1$-$B131 & 25.67882 & 19.02030 & 0.37839 & $ 20.52 \pm0.03 $ & $ 0.28 \pm  0.04 $ & $ 0.87 \pm  0.06 $ & $ 409 \pm  40 $ & $ -20.18 \pm  0.10 $ & $ 0.833 \pm  0.053 $ & $ 2.105 \pm  0.043 $ \\
SFF10$-$NB1$-$D4520 & 25.97511 & 28.02058 & 0.37922 & $ 22.36 \pm0.07 $ & $ -0.38 \pm  0.12 $ & $ 1.08 \pm  0.12 $ & $ 481 \pm  68 $ & $ -18.13 \pm  0.13 $ & $ 0.740 \pm  0.077 $ & $ 1.029 \pm  0.058 $ \\
SFF10$-$NB1$-$A3582 & 26.37815 & 27.65652 & 0.40632 & $ 22.62 \pm0.09 $ & $ 0.29 \pm  0.12 $ & $ 0.49 \pm  0.11 $ & $ 320 \pm  18 $ & $ -18.65 \pm  0.11 $ & $ 0.539 \pm  0.055 $ & $ 0.955 \pm  0.067 $ \\
SFF10$-$NB1$-$A2772 & 26.40089 & 27.64702 & 0.40398 & $ 22.28 \pm0.07 $ & $ 0.80 \pm  0.08 $ & $ 0.76 \pm  0.09 $ & $ 357 \pm  15 $ & $ -18.69 \pm  0.11 $ & $ 0.669 \pm  0.074 $ & $ 1.282 \pm  0.054 $ \\
SFF17$-$NB1$-$C35670 & 26.78536 & 28.12596 & 0.39224 & $ 22.53 \pm0.08 $ & $ 0.26 \pm  0.12 $ & $ 0.27 \pm  0.12 $ & $ 587 \pm  55 $ & $ -18.86 \pm  0.12 $ & $ 0.738 \pm  0.043 $ & $ 1.212 \pm  0.040 $ \\
SFF22$-$NB1$-$B1629 & 30.26609 & 19.87761 & 0.38235 & $ 22.57 \pm0.07 $ & $ -0.05 \pm  0.13 $ & $ 1.00 \pm  0.13 $ & $ 541 \pm  54 $ & $ -18.02 \pm  0.14 $ & $ 0.738 \pm  0.030 $ & $ 1.034 \pm  0.041 $ \\
SFF22$-$NB1$-$A7322 & 30.41768 & 19.96039 & 0.38553 & $ 23.21 \pm 0.10 $ & $ -0.07 \pm  0.20 $ & $ 0.54 \pm  0.15 $ & $ 427 \pm  59 $ & $ -17.86 \pm  0.14 $ & $ 0.600 \pm  0.045 $ & $ 0.599 \pm  0.048 $ \\
SFF22$-$NB1$-$A1201 & 30.59303 & 19.69401 & 0.37849 & $ 22.34 \pm0.06 $ & $ -0.07 \pm  0.10 $ & $ 0.72 \pm  0.10 $ & $ 361 \pm  59 $ & $ -18.50 \pm  0.12 $ & $ 0.784 \pm  0.042 $ & $ 0.969 \pm  0.066 $ \\
SFF14$-$NB1$-$B10370 & 38.35027 & 27.29842 & 0.39459 & $ 22.43 \pm 0.10 $ & $ 0.31 \pm  0.16 $ & $ 0.33 \pm  0.13 $ & $ 413 \pm  43 $ & $ -18.91 \pm  0.12 $ & $ 0.718 \pm  0.029 $ & $ 1.099 \pm  0.043 $ \\
SFF24$-$NB1$-$D2771 & 48.38854 & 19.81745 & 0.40538 & $ 22.76 \pm0.09 $ & $ 0.29 \pm  0.13 $ & $ 0.52 \pm  0.15 $ & $ 305 \pm  20 $ & $ -18.46 \pm  0.15 $ & $ 0.624 \pm  0.042 $ & $ 0.752 \pm  0.072 $ \\
SFS02$-$NB1$-$C26028 & 131.94269 & 13.58708 & 0.38353 & $ 22.48 \pm0.08 $ & $ -0.30 \pm  0.16 $ & $ 0.49 \pm  0.12 $ & $ 308 \pm  37 $ & $ -18.61 \pm  0.12 $ & $ 0.671 \pm  0.053 $ & $ 0.805 \pm  0.047 $ \\
SFS04$-$NB1$-$D7759 & 134.62971 & 5.09669 & 0.38701 & $ 21.61 \pm0.05 $ & $ 0.16 \pm  0.09 $ & $ 0.51 \pm  0.10 $ & $ 473 \pm  41 $ & $ -19.50 \pm  0.12 $ & $ 0.801 \pm  0.015 $ & $ 1.565 \pm  0.038 $ \\
SFS04$-$NB1$-$C37846 & 134.98694 & 5.23336 & 0.38409 & $ 22.61 \pm0.08 $ & $ -0.20 \pm  0.16 $ & $ 0.44 \pm  0.14 $ & $ 405 \pm  44 $ & $ -18.54 \pm  0.15 $ & $ 0.679 \pm  0.024 $ & $ 0.891 \pm  0.043 $ \\
SFS05$-$NB1$-$A5037 & 139.71170 & 13.62559 & 0.39218 & $ 22.74 \pm 0.10 $ & $ 0.16 \pm  0.17 $ & $ 0.99 \pm  0.20 $ & $ 338 \pm  39 $ & $ -17.92 \pm  0.19 $ & $ 0.606 \pm  0.041 $ & $ 0.867 \pm  0.045 $ \\
SFS07$-$NB1$-$D13040 & 185.08850 & 43.59100 & 0.39999 & $ 22.29 \pm0.06 $ & $ 0.09 \pm  0.13 $ & $ 0.71 \pm  0.13 $ & $ 477 \pm  62 $ & $ -18.71 \pm  0.14 $ & $ 0.671 \pm  0.051 $ & $ 1.290 \pm  0.055 $ \\
SFS07$-$NB1$-$D4656 & 185.36786 & 43.35006 & 0.38968 & $ 23.16 \pm0.09 $ & $ 0.20 \pm  0.16 $ & $ 0.29 \pm  0.14 $ & $ 317 \pm  51 $ & $ -18.19 \pm  0.14 $ & $ 0.457 \pm  0.064 $ & $ 0.626 \pm  0.063 $ \\
SFS07$-$NB1$-$D1714 & 185.46849 & 43.45247 & 0.37956 & $ 22.73 \pm0.07 $ & $ 0.04 \pm  0.15 $ & $ 0.61 \pm  0.14 $ & $ 388 \pm  73 $ & $ -18.23 \pm  0.15 $ & $ 0.646 \pm  0.048 $ & $ 0.894 \pm  0.076 $ \\
SFS07$-$NB1$-$C9757 & 185.64703 & 43.22250 & 0.38949 & $ 22.02 \pm0.04 $ & $ 0.23 \pm  0.08 $ & $ 0.67 \pm  0.08 $ & $ 314 \pm  30 $ & $ -18.95 \pm  0.11 $ & $ 0.719 \pm  0.047 $ & $ 1.228 \pm  0.042 $ \\
SFS07$-$NB1$-$C6636 & 185.73920 & 43.29079 & 0.38191 & $ 23.44 \pm0.09 $ & $ -0.02 \pm  0.19 $ & $ 0.98 \pm  0.21 $ & $ 377 \pm  68 $ & $ -17.16 \pm  0.21 $ & $ 0.541 \pm  0.169 $ & $ 0.492 \pm  0.068 $ \\
SFS09$-$NB1$-$D5407 & 198.46715 & 44.11563 & 0.39802 & $ 22.76 \pm0.07 $ & $ 0.38 \pm  0.10 $ & $ 0.49 \pm  0.12 $ & $ 314 \pm  36 $ & $ -18.44 \pm  0.13 $ & $ 0.512 \pm  0.067 $ & $ 0.914 \pm  0.047 $ \\
SFS09$-$NB1$-$A11676 & 198.62309 & 43.58524 & 0.40364 & $ 22.91 \pm0.08 $ & $ 0.19 \pm  0.13 $ & $ 0.76 \pm  0.13 $ & $ 333 \pm  21 $ & $ -18.06 \pm  0.14 $ & $ 0.626 \pm  0.149 $ & $ 0.804 \pm  0.070 $ \\
SFS10$-$NB1$-$D7275 & 198.68433 & 43.83075 & 0.39628 & $ 22.61 \pm 0.10 $ & $ 0.57 \pm  0.13 $ & $ 0.20 \pm  0.13 $ & $ 330 \pm  35 $ & $ -18.86 \pm  0.12 $ & $ 0.769 \pm  0.018 $ & $ 1.048 \pm  0.045 $ \\
SFS09$-$NB1$-$A7620 & 198.75090 & 43.59008 & 0.39627 & $ 22.31 \pm0.05 $ & $ 0.42 \pm  0.08 $ & $ 0.47 \pm  0.08 $ & $ 318 \pm  29 $ & $ -18.90 \pm  0.11 $ & $ 0.723 \pm  0.020 $ & $ 1.154 \pm  0.040 $ \\
SFS10$-$NB1$-$D5176 & 198.75093 & 43.59009 & 0.39646 & $ 22.33 \pm0.06 $ & $ 0.33 \pm  0.08 $ & $ 0.42 \pm  0.09 $ & $ 315 \pm  31 $ & $ -18.93 \pm  0.11 $ & $ 0.703 \pm  0.101 $ & $ 1.120 \pm  0.043 $ \\
SFS10$-$NB1$-$A22133 & 198.92332 & 43.48064 & 0.38810 & $ 22.56 \pm0.07 $ & $ 0.57 \pm  0.09 $ & $ 0.49 \pm  0.10 $ & $ 895 \pm  78 $ & $ -18.57 \pm  0.11 $ & $ 0.824 \pm  0.013 $ & $ 1.474 \pm  0.038 $ \\
SFS10$-$NB1$-$C7170 & 199.18407 & 43.57734 & 0.40692 & $ 22.87 \pm0.09 $ & $ 0.41 \pm  0.11 $ & $ 0.31 \pm  0.11 $ & $ 390 \pm  26 $ & $ -18.57 \pm  0.11 $ & $ 0.759 \pm  0.015 $ & $ 0.980 \pm  0.082 $ \\
\enddata
\tablecomments{This table is available in its entirety in machine-readable form. \added{$a$. KISS GP (\citealt{Brunker2020})}}
\label{tab: NB695}
\end{deluxetable}
\end{longrotatetable}
\begin{longrotatetable}
\begin{deluxetable}{lccccrrcccc}
\tablecaption{NB746 SFACT Green Peas Candidates}
\tablewidth{0pt}
\tabletypesize{\scriptsize}
\tablehead{
\colhead{SFACT Object ID} & \colhead{$\alpha$(J2000)} & \colhead{$\delta$(J2000)} & \colhead{\textit{z}} &
\colhead{$m_{r}$} & \colhead{$r - i$} & \colhead{$g - r$} &
\colhead{EW$_{5007,phot}$} & \colhead{$M_{g}$}& \colhead{log([O III]/H$\beta$)} & \colhead{log(SFR)}
\\
\colhead{ } & \colhead{(deg)} & \colhead{(deg)} & \colhead{} &
\colhead{(mag)} & \colhead{(mag)} & \colhead{(mag)} &
\colhead{($\text{\AA}$)} & \colhead{(mag)}&\colhead{} & \colhead{(M$_{\odot}$ yr$^{-1}$)}
\\
\colhead{(1)} & \colhead{(2)} & \colhead{(3)} & \colhead{(4)} &
\colhead{(5)} & \colhead{(6)} & \colhead{(7)} &
\colhead{(8)} & \colhead{(9)}&\colhead{(10)}& \colhead{(11)}
} 
\startdata
SFF09$-$NB3$-$B3508 & 20.80180 & 28.02868 & 0.49059 & $ 22.21 \pm0.05 $ & $ 0.47 \pm  0.08 $ & $ 0.34 \pm  0.08 $ & $ 338 \pm  21 $ & $ -19.70 \pm  0.10 $ & $ 0.629 \pm  0.148 $ & $ 1.571 \pm  0.041 $ \\
SFF22$-$NB3$-$A7229 & 30.42048 & 19.65701 & 0.48151 & $ 22.07 \pm0.09 $ & $ 0.87 \pm  0.11 $ & $ -0.06 \pm  0.11 $ & $ 399 \pm  22 $ & $ -20.17 \pm  0.11 $ & $ 0.718 \pm  0.015 $ & $ 1.839 \pm  0.039 $ \\
SFF18$-$NB3$-$A13478 & 42.09252 & 20.31637 & 0.48895 & $ 22.24 \pm0.08 $ & $ 0.83 \pm  0.10 $ & $ 0.28 \pm  0.12 $ & $ 415 \pm  22 $ & $ -19.72 \pm  0.13 $ & $ 0.733 \pm  0.098 $ & $ 1.732 \pm  0.039 $ \\
SFF24$-$NB3$-$D12712 & 48.15273 & 19.79708 & 0.50964 & $ 22.19 \pm0.04 $ & $ 0.59 \pm  0.06 $ & $ 0.21 \pm  0.06 $ & $ 713 \pm  16 $ & $ -19.94 \pm  0.10 $ & $ 0.725 \pm  0.145 $ & $ 1.930 \pm  0.043 $ \\
SFF24$-$NB3$-$B29834 & 48.28959 & 19.26982 & 0.50781 & $ 22.76 \pm0.09 $ & $ 0.77 \pm  0.11 $ & $ -0.16 \pm  0.11 $ & $ 423 \pm  17 $ & $ -19.73 \pm  0.11 $ & $ 0.408 \pm  0.058 $ & $ 1.468 \pm  0.058 $ \\
SFS02$-$NB3$-$D4010 & 131.78246 & 13.73971 & 0.49355 & $ 22.44 \pm0.10 $ & $ 0.75 \pm  0.12 $ & $ 0.40 \pm  0.13 $ & $ 390 \pm  21 $ & $ -19.40 \pm  0.12 $ & $ 0.679 \pm  0.021 $ & $ 1.635 \pm  0.039 $ \\
SFS07$-$NB3$-$A3198 & 185.86623 & 43.10892 & 0.50675 & $ 22.70 \pm0.06 $ & $ 0.82 \pm  0.08 $ & $ 0.62 \pm  0.10 $ & $ 333 \pm  11 $ & $ -19.01 \pm  0.12 $ & $ 0.698 \pm  0.122 $ & $ 1.533 \pm  0.050 $ \\
SFS11$-$NB3$-$B22598 & 214.78731 & 43.49066 & 0.49294 & $ 22.70 \pm0.09 $ & $ 0.49 \pm  0.12 $ & $ 0.41 \pm  0.14 $ & $ 433 \pm  26 $ & $ -19.14 \pm  0.13 $ & $ 0.614 \pm  0.036 $ & $ 1.490 \pm  0.040 $ \\
SFS14$-$NB3$-$B11054 & 235.34511 & 43.79404 & 0.48107 & $ 22.07 \pm0.06 $ & $ 0.27 \pm  0.09 $ & $ 0.65 \pm  0.09 $ & $ 311 \pm  28 $ & $ -19.46 \pm  0.11 $ & $ 0.814 \pm  0.039 $ & $ 1.547 \pm  0.049 $ \\
SFS14$-$NB3$-$D3878 & 235.49232 & 44.01439 & 0.48059 & $ 22.91 \pm0.08 $ & $ 0.34 \pm  0.13 $ & $ 0.79 \pm  0.14 $ & $ 319 \pm  48 $ & $ -18.49 \pm  0.14 $ & $ 0.644 \pm  0.051 $ & $ 1.149 \pm  0.069 $ \\
SFS17$-$NB3$-$A8343 & 236.17024 & 29.18409 & 0.48058 & $ 21.49 \pm0.04 $ & $ 0.63 \pm  0.07 $ & $ 0.45 \pm  0.06 $ & $ 359 \pm  21 $ & $ -20.24 \pm  0.10 $ & $ 0.749 \pm  0.009 $ & $ 2.049 \pm  0.041 $ \\
SFF01$-$NB3$-$A13758 & 325.86777 & 19.99045 & 0.48413 & $ 23.00 \pm 0.10 $ & $ 0.89 \pm  0.13 $ & $ 0.23 \pm  0.14 $ & $ 379 \pm  25 $ & $ -18.98 \pm  0.13 $ & $ 0.780 \pm  0.021 $ & $ 1.341 \pm  0.041 $ \\
SFF02$-$NB3$-$C35963 & 330.00958 & 20.06305 & 0.50669 & $ 22.47 \pm0.09 $ & $ 0.40 \pm  0.13 $ & $ 0.35 \pm  0.13 $ & $ 366 \pm  25 $ & $ -19.51 \pm  0.13 $ & $ 0.581 \pm  0.081 $ & $ 1.474 \pm  0.087 $ \\
SFF19$-$NB3$-$D20721 & 337.11969 & 20.08900 & 0.49240 & $ 23.23 \pm 0.10 $ & $ 0.65 \pm  0.14 $ & $ 0.68 \pm  0.17 $ & $ 446 \pm  33 $ & $ -18.35 \pm  0.16 $ & $ 0.708 \pm  0.075 $ & $ 1.267 \pm  0.043 $ \\
SFF06$-$NB3$-$A6250 & 350.14401 & 24.83801 & 0.49111 & $ 23.24 \pm0.09 $ & $ 0.65 \pm  0.13 $ & $ 0.75 \pm  0.17 $ & $ 414 \pm  32 $ & $ -18.26 \pm  0.16 $ & $ 0.831 \pm  0.044 $ & $ 1.216 \pm  0.044 $ \\
\enddata
\tablecomments{This table is available in machine-readable form.}
\label{tab: NB746}
\end{deluxetable}
\end{longrotatetable}

\noindent of the SFACT GP candidates and compare them to other GP samples in Section \ref{subsec:SFR}. We note that our sample \noindent spans three redshift windows, each with a different selection function. 

We catalog the observed and derived properties of the [\ion{O}{3}]-detected SFACT GP candidates, in Table \ref{tab: NB659}, Table \ref{tab: NB695}, and Table \ref{tab: NB746}. In each Table, column (1) lists the SFACT Object ID. This is a unique identifier for each SFACT source (\citetalias{Salzer2023}; \citetalias{Sieben2023}; \citetalias{Carr2023}). The identifier consists of the field designation (ex: SFF07), the filter in which the source was detected (ex: NB2), and the quadrant of the ODI images in which the source was found (ex: A), and a running number assigned during the initial object-detection stage (ex: A7587). We note that this work utilizes different nomenclature for the narrow-band filters (see Section \ref{sec:data}), but the official SFACT names utilize the filter nomenclature presented in \citetalias{Salzer2023}, \citetalias{Sieben2023}, and \citetalias{Carr2023}.

Columns (2) and (3) list the J2000 coordinates of the SFACT GP candidates. Columns (4) lists the redshifts measured from the Hydra spectra of each galaxy. The \textit{r} magnitudes, $r - i$ color, and $g - r$ color measured from the ODI images for each galaxy are listed in columns (5), (6), and (7). Each of these values have been corrected for Galactic extinction using the data from \citet{schlafly2011}. In column (8), we present the EW$_{5007,phot}$ values for each galaxy, derived from the NB and BB photometry as describe in Section \ref{sec:O3_det}. Column (9) presents the g-band absolute magnitudes. These are observed luminosities and have not been K-corrected. Column (10) lists the log([\ion{O}{3}]/H$\beta$) in Column (10). We also list the star formation rates (SFRs) in Column (11); these will be discussed further in Section \ref{subsec:SFR}. 

Of the 123 GP candidates identified using our selection criteria, 11 were previously known to be GPs from either the KISS (\citetalias{Brunker2020}) or H$\alpha$ Dot (\citetalias{Kimsey-Miller2024}) samples.  As stressed in \citetalias{Salzer2023}, many of the SFACT survey fields are centered upon previously known KISS and H$\alpha$ Dot GPs.  In most cases where the redshift of the KISS or H$\alpha$ Dot GP places their [\ion{O}{3}] lines within the bandpass of one of the SFACT NB filters and their BB colors and EWs satisfy our criteria, they have been redetected as SFACT GP candidates.  Of the remaining 112 SFACT GP candidates, four NB2 GP candidates were also cataloged as GP candidates without spectra in \citet{Samonski2025}. We also note that one SFACT NB3 GP candidate, SFS17$-$NB3$-$A8343, has a spectrum in DESI DR1 \citep{desi2026}. The remaining 107 SFACT GP candidates are typically fainter than any of the previously known \citetalias{Cardamone2009}, \citetalias{Brunker2020}, or \citetalias{Kimsey-Miller2024} GPs, exhibiting a range of g-band apparent magnitudes of 21.2 to 24.5 (median value \textit{g} = 23.0). Hence, with only a few exceptions (noted in the Table notes), most SFACT GP candidates are previously unknown systems.

\begin{figure*}[t!]
\centering
\includegraphics[width=0.8\linewidth]{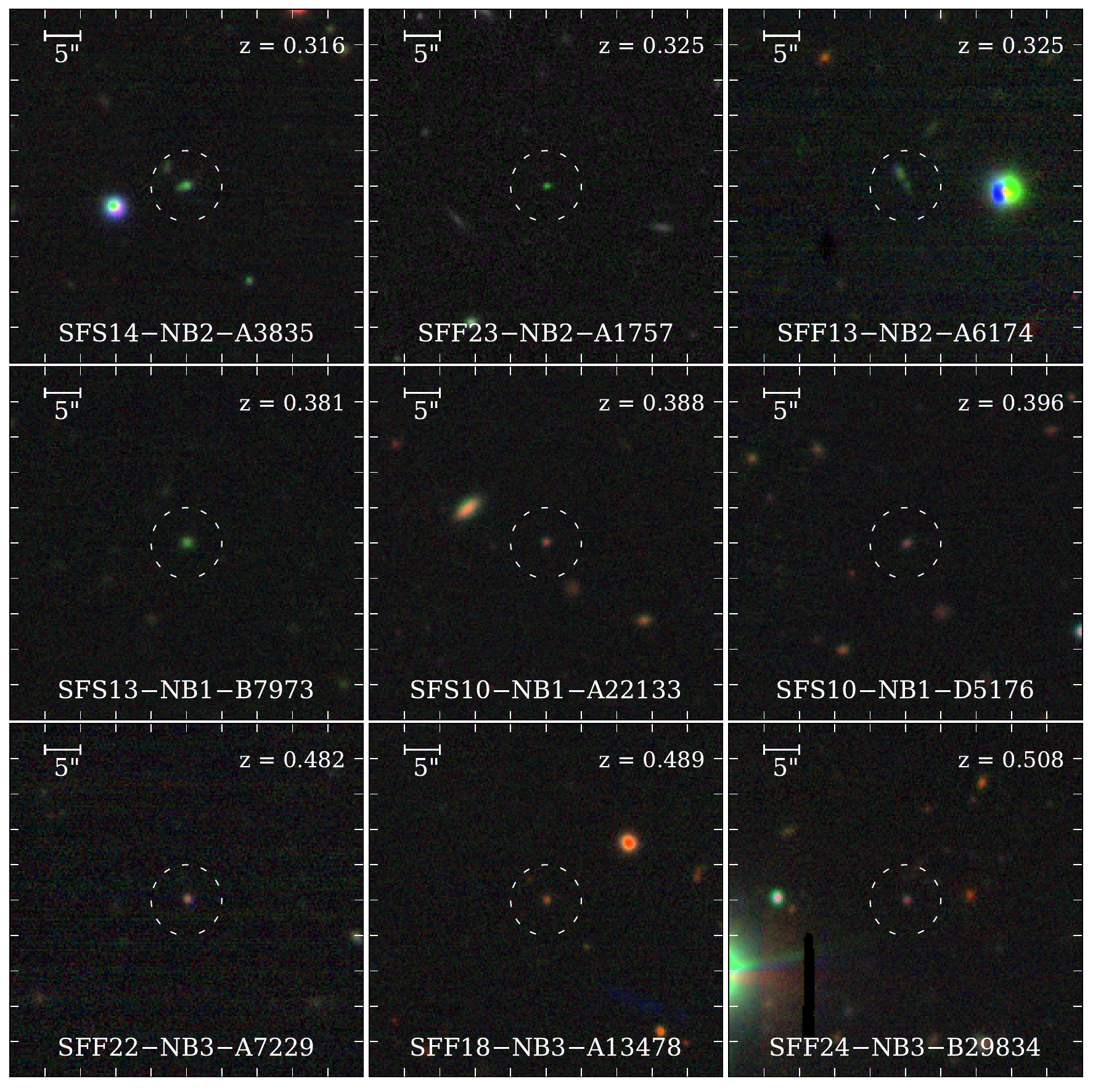} 
\caption{Color-composite images of three SFACT NB659 GP candidates (top), NB695 GP candidates (middle), and NB746 GP candidates (bottom) derived using ODI \textit{g}, \textit{r}, and \textit{i} images. The mapping of these images mimic SDSS three-color images, following \citep{Lupton2004}, where ODI \textit{g} is mapped to blue, ODI \textit{r} is mapped to green, and ODI \textit{i} is mapped red. Each image shows 50$\arcsec$~$\times$~50$\arcsec$ cutouts. The white circle represents a 5$\arcsec$ radius. }
\label{fig:gri_grid}
\end{figure*}

\begin{figure*}[t!]
\centering
\includegraphics[width=0.9\linewidth]{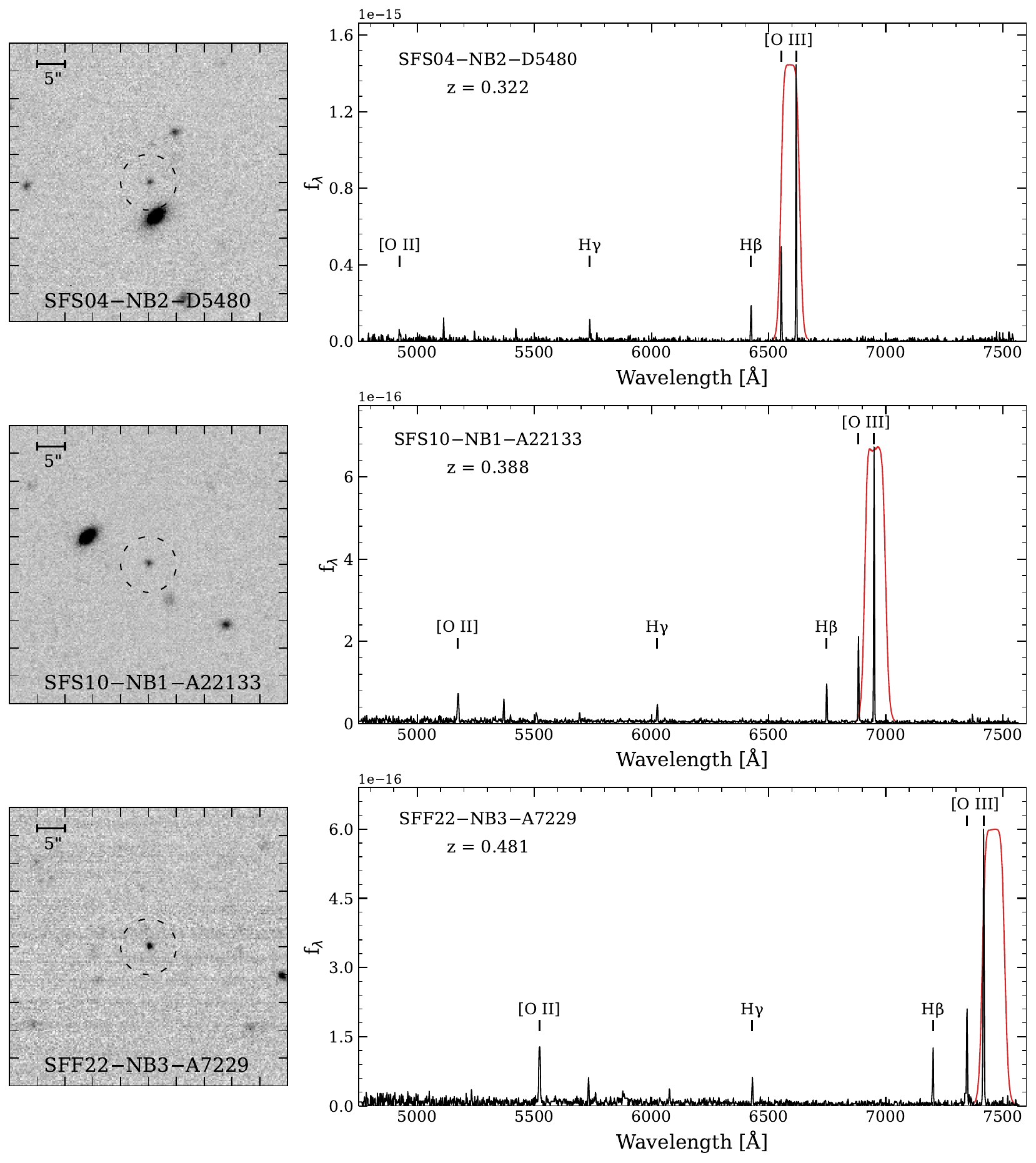} 
\caption{{Left:} ODI \textit{r}-band images for an SFACT NB659 GP candidate (top), an SFACT NB695 GP candidate (middle), and an SFACT NB746 GP candidate (bottom). Each image shows 50$\arcsec$ $\times$ 50$\arcsec$ cutouts, and the black circle represents a 5$\arcsec$ radius. \textit{Right:} The corresponding spectrum for the each GP candidate. The NB filter tracings for a given detection are represented in red. Key emission lines are labeled. The spectra exhibit the characteristic strong [\ion{O}{3}] $\lambda\lambda$ 4959, 5007 emission present in GPs in addition to several other strong emission lines. } 
\label{fig:all_gri_spec_comp_v2}
\end{figure*}

\section{Discussion: Properties of SFACT Green Peas} \label{sec:prop}
In the current section, we explore the properties and characteristics of the SFACT GP candidates. We aim to place them into context with other known samples of GPs. We utilize data from \citetalias{Cardamone2009}, \citetalias{Brunker2020}, and \citetalias{Kimsey-Miller2024} for our comparisons. We note that the full \citetalias{Cardamone2009} GP samples is composed of both SFGs and AGNs. However, for this work we will only consider the 80 \citetalias{Cardamone2009} sources confirmed to be SFGs. 

\subsection{Appearance in SFACT ODI Images}\label{subsec:appearance}
We begin by comparing the physical appearance and spectra of our sample of GP candidates with known GPs. We stress that the appearance of GPs in the three color images is a result of the placement of the [\ion{O}{3}] $\lambda$5007 line and therefore the redshift. The effect of [\ion{O}{3}] emission on the appearance of GP-like objects in three-color images is demonstrated in Figure \ref{fig:gri_grid}. We present \textit{gri} three-color images of nine SFACT GP candidates created using images obtained with ODI. We mimic the color mapping associated with the SDSS three-color images where \textit{g} is mapped to blue, \textit{r} is mapped to green, and \textit{i} is mapped to red. SFACT NB659 GP candidates (top row) have the classic green appearance associated with \citetalias{Cardamone2009} GPs. 

The middle row in Figure \ref{fig:gri_grid} represents the NB695 GP candidates. As previously mentioned, at z~$\gtrsim$~0.36 GPs appear purple in the SDSS three-color images and gain the designation of ``Purple Grapes" (\citetalias{Brunker2020}; \citealt{Liu2022}). However, we notice that NB695 GP candidates towards the lower redshift end have similar appearances to the NB659 GP candidate sample, as demonstrated by SFS13$-$NB1$-$B7973 in Figure \ref{fig:gri_grid}. The other NB695 GP candidates presented in Figure \ref{fig:gri_grid}, SFS10$-$NB1$-$A22133 and SFS10$-$NB1$-$D5176, have similar appearances to the Purple Grapes. We attribute this effect on appearance to differences in the SDSS and ODI filter bandpasses. At the redshift of SFS13$-$NB1$-$B7973, z~=~0.381, the [\ion{O}{3}]$\lambda$5007 line falls at a wavelength where the transmission in the ODI \textit{r} is $\sim$50$\%$. This is sufficient transmission of [\ion{O}{3}]$\lambda$5007 in the ODI \textit{r} to cause the green appearance of SFS13$-$NB1$-$B7973 in the ODI three-color images. We note the SFS13$-$NB1$-$B7973 does have the expected purple hue in the SDSS three-color images. By the redshift of SFS10$-$NB1$-$A22133, z~=~0.388, the [\ion{O}{3}]$\lambda$5007 has sufficiently redshifted enough into the ODI \textit{i}, and the NB695 GP candidates no longer appear green. 

We also present NB746 GP candidates in the bottom row of Figure \ref{fig:gri_grid}. In this redshift range, z~=~0.48~--~0.51, the [\ion{O}{3}]$\lambda$5007 line has fully redshifted into the ODI \textit{i} filter, boosting the \textit{i} flux. As expected, these objects have purple or red hues and appear similar to the higher redshift NB695 GP candidates.

The SFACT images have a median seeing of 0$\farcs$8, compared to the seeing of 1$\farcs$3 in SDSS images. At this resolution, most of our sample is unresolved or marginally resolved. The resolution limit of the SFACT images corresponds to an upper limit on the physical radius for unresolved GP candidates. At the redshifts of the SFACT GP candidates, the angular scales correspond to physical radii of $\sim$3~kpc, $\sim$4~kpc, and $\sim$6~kpc for NB659, NB695, and NB746 GP candidates, respectively. It is notable that despite not applying a morphological criterion when selecting our sample, our criteria preferentially select objects that appear compact in our images. 

While SFACT images are insufficient to spatially resolve SFACT GP candidates, the improved resolution over SDSS allows us to identify multiple systems that are unresolved in SDSS. A notable example of this is SFF13$-$NB2$-$A6174 presented in top right corner of Figure \ref{fig:gri_grid}, which resembles the ``Extended Peas" found in \citet{Samonski2025}. In SDSS, this system appears as a single amorphous extended object, whereas the higher resolution afforded by ODI reveals multiple galaxy components within the 5$\arcsec$ circle. The second galaxy component visible, SFF13$-$NB2$-$A6161, is a distinct SFACT NB659 [\ion{O}{3}] detection in the immediate environment of SFF13$-$NB2$-$A6174, with a projected separation of $\sim$400~kpc. We note that SFF13$-$NB2$-$A6161 falls just below our EW$_{5007,phot}$ cut with an EW$_{5007,phot}$ of 277~$\pm$~31~$\text{\AA}$, but otherwise satisfies our other selection criteria for NB659 GP candidates (e.g., $r - i \leq -0.1$ and 0.4~$\leq$~log([\ion{O}{3}]/H$\beta$ $\leq$~0.9).

The presence of two extreme, GP-like systems in close proximity is particularly interesting. Previous large-scale environment studies have found that GPs primarily reside in low density environments on (\citetalias{Cardamone2009}; \citealt{Brunker2022}; \citealt{Liu2022}; \citealt{Ding2023}; \citetalias{Kimsey-Miller2024}; \citealt{Gupta2026}). A study of the large-scale environment of a subsample of SFACT GP candidates in z~=~0.31~--~0.49 is presented in \citet{KM26}, who find that the majority of SFACT GP candidates in their subsample reside in extremely low-density environments. Recent results from \citet{Arroyo-Polonio2026} have shown that on smaller scales, GPs reside in local overdensities of ELG companions. While galaxy mergers and interactions can trigger star formation (e.g., \citealt{Barton2000};\citealt{Lambas2003}; \citealt{Brosch2004}; \citealt{Barton2007}; \citealt{Li2008}), they are not necessary to drive activity in GPs (\citealt{Amor2015}; \citealt{Lofthouse2017}; \citealt{Laufman2022}; \citealt{Arroyo-Polonio2023}; \citealt{Brunker2022}; \citetalias{Kimsey-Miller2024}; \citealt{Arroyo-Polonio2026}; \citealt{Gupta2026}).

\begin{figure}[t!]
\centering
\includegraphics[width=\linewidth]{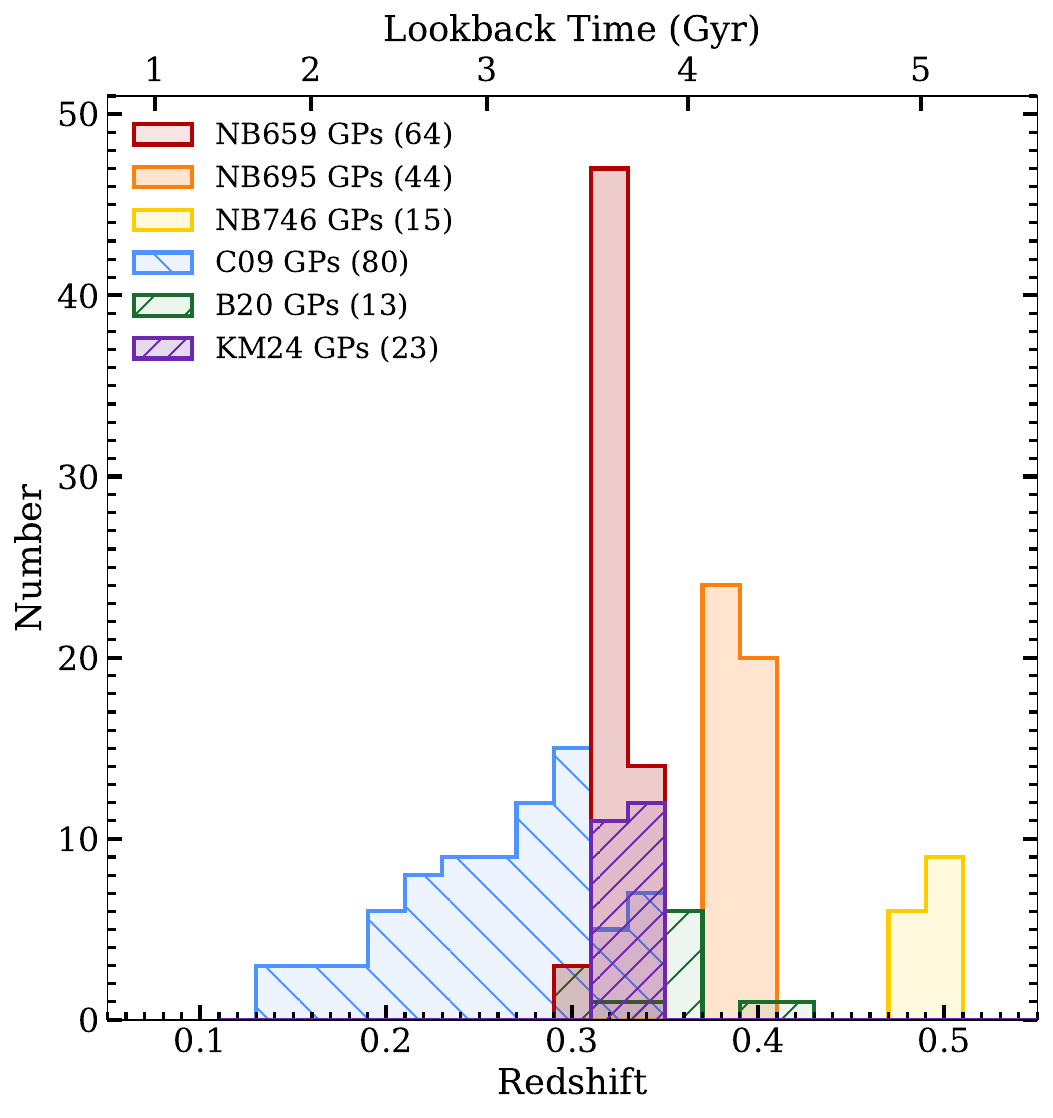} 
\caption{Histogram of the redshift distribution of the NB659 (red), NB695 (orange), NB746 (yellow) GP candidate samples from SFACT (this work), \citet{Cardamone2009} (blue), \citet{Brunker2020} (green), and \citet{Kimsey-Miller2024} (purple). There is significant overlap in redshift in the NB659 redshift window, z~=~0.31~--~0.34. Meanwhile, there is slight or no overlap in the NB695 and NB746 redshift windows, z~=~0.38~--~0.41 and z~=~0.48~--~0.51, respectively. The SFACT sample is the dominant sample in each redshift window.}
\label{fig:gp_red_dist}
\end{figure}

The characteristic spectral signature of the SFACT GP candidates is demonstrated in Figure \ref{fig:all_gri_spec_comp_v2} in which we present spectra of one SFACT GP candidate in each filter. The SFACT candidates presented in Figure \ref{fig:all_gri_spec_comp_v2} have spectra reminiscent of \citetalias{Cardamone2009} and \citetalias{Brunker2020} GPs, possessing characteristic high-excitation spectra and strong [\ion{O}{3}]$\lambda$$\lambda$4959,5007 emission lines associated with GPs. SFS04$-$NB2$-$D5480 exhibits some of the most extreme spectral properties in our sample. Despite being among the least luminous in our sample (M$_g=-17.40$), SFS04$-$NB2$-$D5480 exhibits the strongest EW$_{5007}$ of 1760~$\pm$~202~$\text{\AA}$, measurable [\ion{O}{3}] $\lambda$4363, and very weak [\ion{O}{2}]$\lambda$3727. These properties, particularly the high [\ion{O}{3}]/[\ion{O}{2}] ratio and elevated EW$_{5007,phot}$, are characteristic of low-redshift Lyman continuum (LyC) emitters \citep{Jaskot2013,Nakajima2014,Izotov2016Natur,Izotov2016MNRAS,Izotov2018}. Though a high [\ion{O}{3}]/[\ion{O}{2}] ratio alone is not sufficient enough to determine LyC emitter status \citep{Izotov2017,flury2022}, it signals that SFS04$-$NB2$-$D5480 may be a worthy target for UV follow-up.

\subsection{Redshift Distribution of SFACT GP Candidates} \label{subsec:red_dist}
In Figure \ref{fig:gp_red_dist}, we compare the redshift distributions of the \citetalias{Cardamone2009}, \citetalias{Brunker2020}, \citetalias{Kimsey-Miller2024}, and SFACT GP candidate samples. The different redshift ranges covered by these GP samples are the result of the width and placement of the respective filters used to detect these galaxies. The \citetalias{Cardamone2009} sample covers a redshift range of z~=~0.14~--~0.35, corresponding to the [\ion{O}{3}] $\lambda$5007 line falling in the $>$50$\%$ throughput range of the SDSS \textit{r} filter. Outside of this redshift range, GP-like objects no longer appear green in the SDSS three-color images and are not selected as GPs in the \citetalias{Cardamone2009} sample. The \citetalias{Brunker2020} sample has a redshift range of z~=~0.29~--~0.42, a result of the [\ion{O}{3}] line falling in the wavelength coverage of the KISS red filter and objective prism. The \citetalias{Kimsey-Miller2024} sample is comprised of strong-line sources detected by their strong [\ion{O}{3}] lines in a NB filter. The wavelength coverage and placement of the H$\alpha$ Dot NB filter leads to a redshift range of z =~0.32~--~0.34. As previously discussed, and further illustrated in Figure \ref{fig:gp_red_dist}, there is some overlap in redshift between the comparison GP samples and the SFACT GP candidate sample. However, the SFACT GP candidate sample extends to higher redshifts than the comparison samples. We emphasize that this is due to the placement of the SFACT NB filters allowing for the detections of higher redshift ELGs than these other samples. In each SFACT redshift window the SFACT sample has more GP candidates than the comparison samples.

\subsection{Low-Luminosity NB659 GP Candidates} \label{subsec:low_luminosity}

When taking into account the smaller areal sky coverage of SFACT compared to these other surveys, the fact that we find more GPs in the narrow SFACT redshift windows, particularly in the NB659 window, raises some interesting questions. For example, are GPs more common than expected? To address this question, we explore the areal densities of NB659 SFACT GP candidates compared to GPs in the same overlapping redshift window of the \citetalias{Cardamone2009}, \citetalias{Brunker2020}, and \citetalias{Kimsey-Miller2024} samples. Comparing the densities in the same redshift window removes any effects due to possible cosmic evolution in the volume density of GPs. We note that the \citetalias{Cardamone2009} catalog is drawn from a sample of galaxies with previously available SDSS spectroscopy and is incomplete with respect to the true GP population \citep{Samonski2025}. Therefore, to compare the surface densities of SFACT and SDSS GPs, we must find the expected number of SDSS GPs in the NB659 redshift window. 

\begin{table}
\begin{center}
\caption{Surface Density of GPs in z = 0.31 - 0.34}
    \begin{tabular}{ccll} 
    \hline\hline 
    Survey & No. of GPs & Sky Coverage & Density \\
    & & (deg$^{2}$) & (deg$^{2}$)\\
    \hline
    SFACT & 64 & 21.73 &  2.95 \\
    H$\alpha$ Dots & 23 & 34.42 $^{a,b}$ &  0.67 \\
    KISS & 10 & 128.03 $^{c,d}$ & 0.08 \\
    SDSS DR7 & 1440 & 4900 $^{e}$ & 0.29\\
    \hline 
    \end{tabular}
\end{center}
\tablecomments{Superscripts denote references. $a$.  \citealt{Kellar2012} $b.$ \citealt{Salzer2020} $c$. \citealt{Salzer2001} $d.$ \citealt{Gronwall2004a}  $e$. \citealt{Cardamone2009} as corrected by \citealt{Samonski2025}}
\label{tab:density}
\end{table}\label{tab:density}

\citet{Samonski2025} develops a photometric method to assemble a more complete catalog of GP candidates in SDSS DR18 \citep{Almeida2023}. However, most of their sample has not been observed spectroscopically, and thus have no spectroscopic redshifts or emission-line ratios to confirm the source of ionizing photons (e.g., SFG or AGN). To estimate the total number of SDSS GPs in the NB659 redshift window, we take the fraction of star-forming \citetalias{Cardamone2009} GPs in the NB659 redshift window to total number of \citetalias{Cardamone2009} star-forming GPs and then multiply by the total number of GP candidates in the \citet{Samonski2025} sample ($\sim$9,600). This yields 1440 GPs expected in the redshift window of z~=~0.31~--~0.34 detectable by SDSS. In Table \ref{tab:density}, we list the surface densities of GPs in that redshift window discovered by each sample. SFACT finds 2.95 GPs per sq. deg in this redshift window. This is $\sim$4 times greater than H$\alpha$ Dot GPs, $\sim$10 times greater than SDSS GPs, and $\sim$40 times greater than KISS. We attribute this to improved depth of SFACT compared to these other surveys, allowing us to probe lower luminosity sources.


\begin{figure}[t!]
\centering
\includegraphics[width=\linewidth]{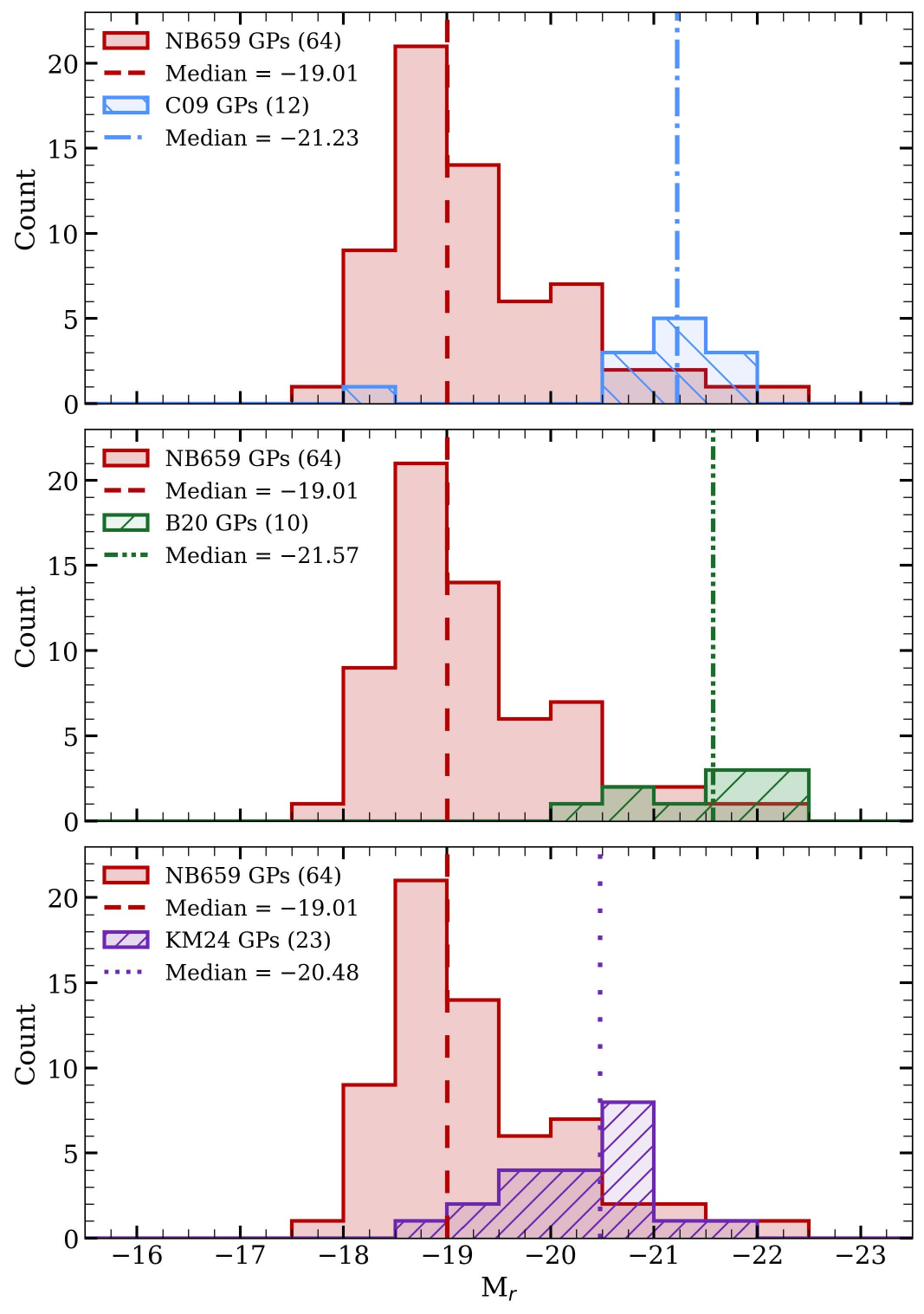} 
\caption{Histogram of the r-band absolute magnitudes of the NB659 GP sample. The r-band absolute magnitudes of 12 \citet[][top panel]{Cardamone2009}, 10 \citet[][middle panel]{Brunker2020}, and 23 \citet[][bottom panel]{Kimsey-Miller2024} GPs in the overlapping redshift window, z~=~0.31~--~0.34 are overplotted. The plot has the same color-coding as in Figure \ref{fig:gp_red_dist}. In all cases, the SFACT NB659 GP consists of lower-luminosity sources with a median r-band absolute magnitude of -19.01 (red dashed line).} 
\label{fig:gp_Mr_comp}
\end{figure}

\begin{figure}[t!]
\centering
\includegraphics[width=\linewidth]{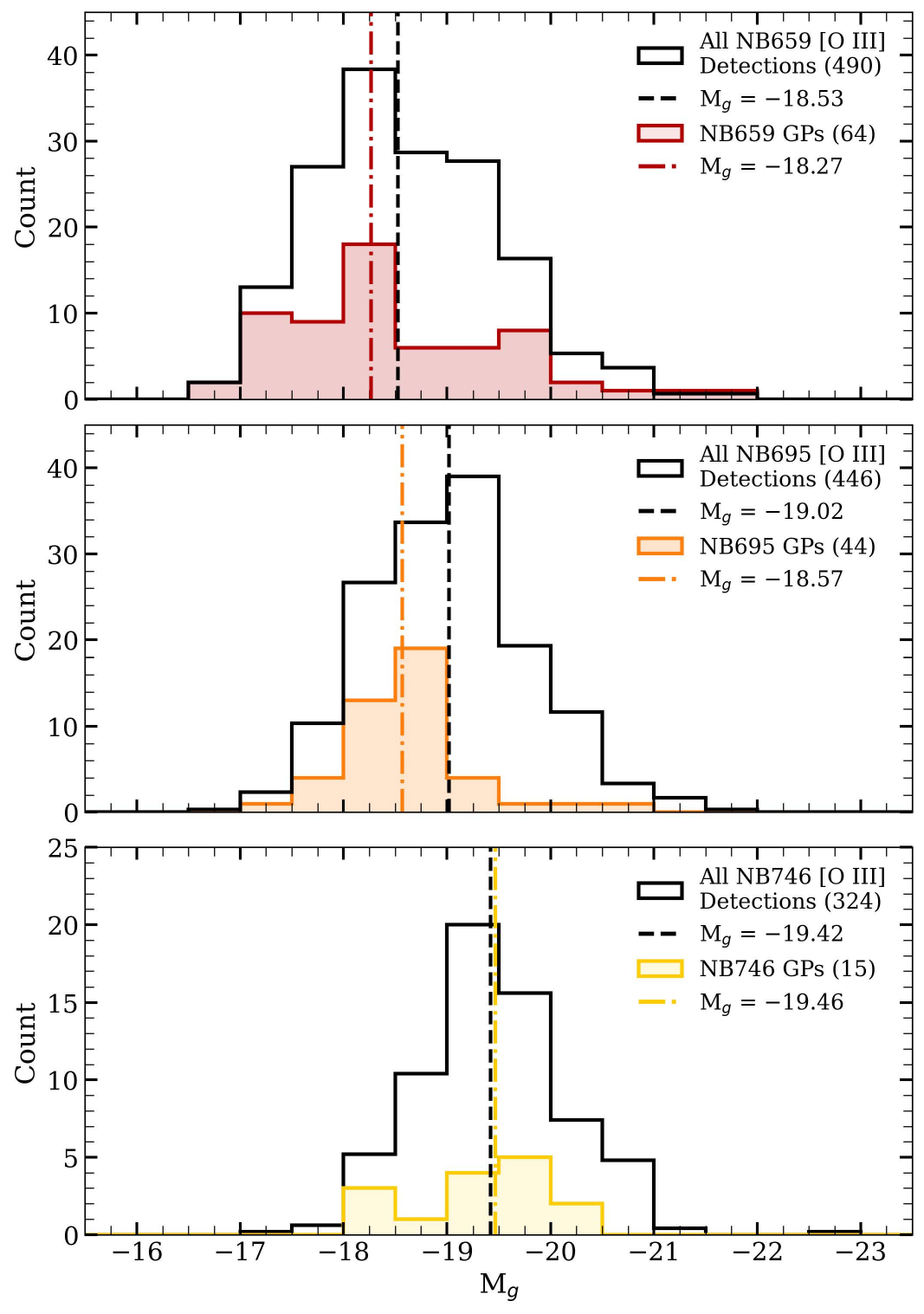} 

\caption{Histograms showing the g-band absolute magnitudes distributions of the NB659 (top panel), NB695 (middle panel), and NB746 (bottom panel) GP samples. The plot has the same color-coding as in Figure \ref{fig:gp_red_dist}. The black histograms represent the full [\ion{O}{3}] sample for each respective NB filter. The black histograms are scaled down to allow for better comparison. The NB659 and NB695 histograms are reduced by a factor of 3, and the NB746 histogram is reduced by a factor of 5. }
\label{fig:mg_filter_comp}
\end{figure}
In Figure \ref{fig:gp_Mr_comp}, we show the distribution of the r-band absolute magnitudes of the NB659 SFACT GP candidates compared to GPs in the NB659 redshift window from the other samples. The NB659 sample spans a wider $M_{r}$ range than the comparison samples, with $-18\gtrsim M_{r}\gtrsim -22$. The NB659 sample possesses a median r-band absolute magnitude of $M_{r} = -19.01$. This is  1.5 magnitudes fainter than the \citetalias{Kimsey-Miller2024} sample, 2.2 magnitudes fainter than the \citetalias{Cardamone2009} sample, and 2.6 magnitudes fainter than the \citetalias{Brunker2020}. Clearly, the SFACT sample is able to probe lower luminosity compared to previous GPs studies in the same redshift window. We note that \citet{Amor2015} also found a population of low luminosity GP-like EELGs. These sources may represent higher redshift Blueberries, which are thought to be low-redshift GP analogs with lower masses and luminosities \citep{Yang2017...blueberry}. Despite being $\sim$2 orders of magnitude less massive, Blueberries have been shown to have sSFRs comparable to the most extreme \citetalias{Cardamone2009} and \citetalias{Brunker2020} GPs \citep{Hirschauer2022,Brunker2022}. However, due to differences in selection methods, it is unclear whether GPs and Blueberries form a continuum of extreme objects or if they are distinct classes of galaxies. Future studies of the NB659 GP candidates may provide insight into the matter.   

\subsection{Absolute Magnitudes of SFACT GP Candidates}
Across the three redshift windows, the size of the SFACT GP sample drops off from 64 at z~$\approx$~0.32 to 15 at z~$\approx$~0.50. To understand why, we present histograms of the g-band absolute magnitudes for the NB659 (top panel), NB695 (middle panel), and NB746 (bottom panels) GP samples in Figure \ref{fig:mg_filter_comp}. We also show the distributions for the full SFACT [\ion{O}{3}]-detected samples in each respective NB filter. The full NB659 and NB695 [\ion{O}{3}]-detected samples cover $-16.5 \gtrsim M_{g} \gtrsim -22$, while the full NB746 [\ion{O}{3}]-detected sample covers $-17 \gtrsim M_{g} \gtrsim -21.5$. The SFACT GP samples generally cover similar luminosity regimes as the full [\ion{O}{3}]-detected populations, though the NB659 and NB695 GP samples have a higher proportion of lower-luminosity sources. The full NB659, NB695, and NB746 [\ion{O}{3}]-detected samples have median $M_{g}$ values of $-18.58$, $-19.02$, and $-19.42$, respectively. The corresponding GP samples have median $M_{g}$ values of $-18.27$, $-18.57$, and $-19.46$. As expected for flux-limited samples, the sensitivity to lower luminosity sources decreases with increasing redshift causing the full GP candidate sample size to drop off with increasing redshift (e.g., there are fewer EW$_{5007}$ $>$ 300~$\text{\AA}$ sources detected with increasing redshift). This is likely the result of a dual metallicity-luminosity effect. Higher luminosity sources tend be metal-rich, resulting in weaker [\ion{O}{3}] emission. These higher luminosity sources also tend to have stronger underlying continuum. These two factors together decrease the strength of the EW$_{5007}$, leading to fewer strong-line sources detectable at high redshifts. This effect is clearly seen in the EW$_{5007,phot}$ distributions shown in Figures \ref{fig:NB659_gr_vsri_EW_subplots} -- \ref{fig:NB746_gr_vsri_EW_subplots}. It is currently not well understood what role the cosmic evolution in GP volume density plays in the decreasing sample size. Our group hopes to explore this issue once a more complete SFACT GP candidate sample is available. 


\subsection{Star Formation Properties of SFACT GP candidates}\label{subsec:SFR}

\begin{figure}[t!]
\centering
\includegraphics[width=\linewidth]{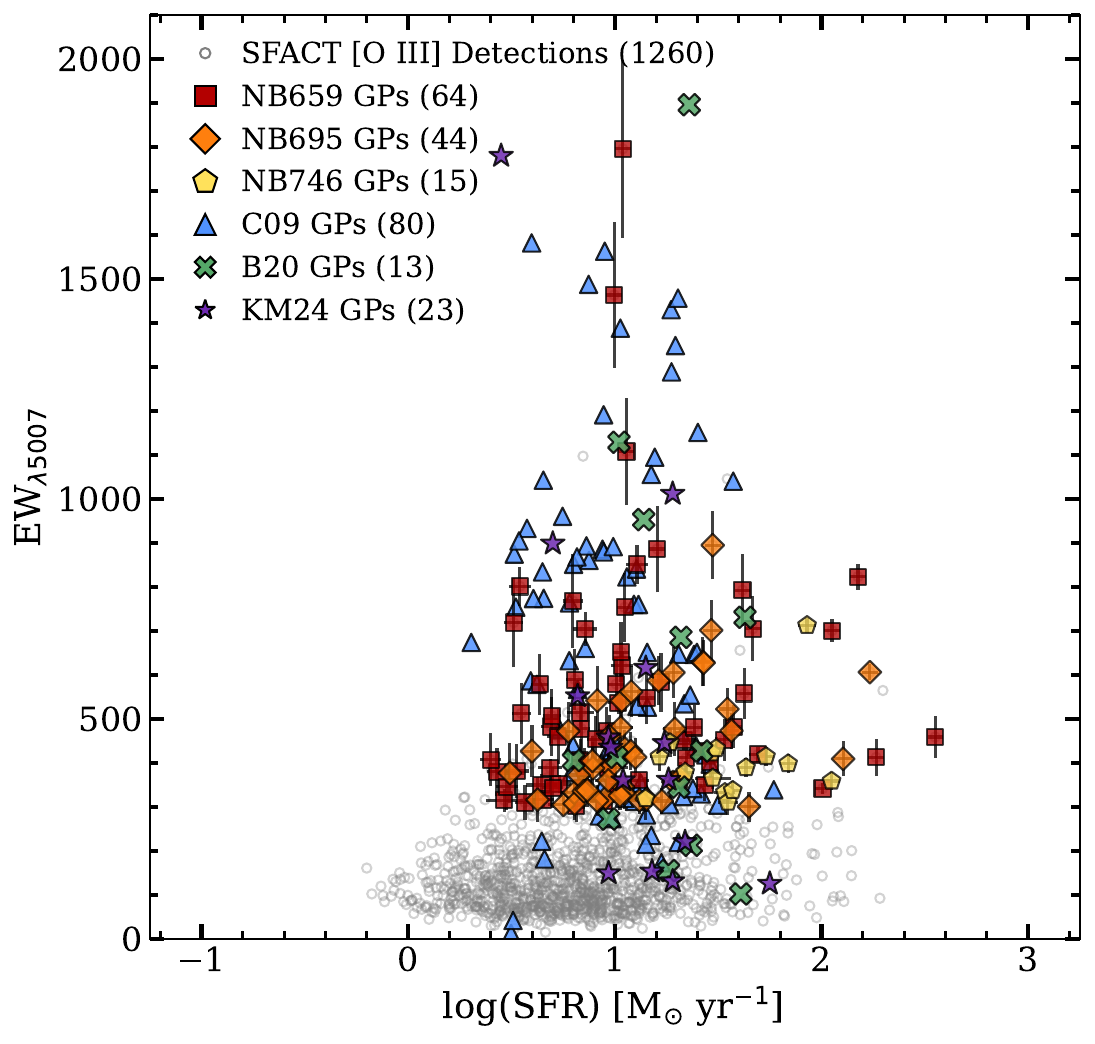} 
\caption{Photometric EW$_{5007,phot}$ plotted against the SFR (derived from [\ion{O}{3}] $\lambda$5007 fluxes) for the SFACT NB659 GP candidates (red squares), NB695 GP candidates (orange diamonds), NB746 GP candidates (yellow pentagons), and full [\ion{O}{3}]-detected sample (grey cricle). GPs from \citet[][blue triangles]{Cardamone2009}, \citet[][green X]{Brunker2020}, and \citet[][purple stars]{Kimsey-Miller2024} are plotted for comparison. The SFACT [\ion{O}{3}]-detected galaxies have some overlap with SFACT GP candidate samples, but the GP candidates tend to be higher SFR sources. The SFACT GP candidates have high SFRS and EW$_{5007}$, occupying the same region as the comparison GP samples. }
\label{fig:EW_SFR_compare}
\end{figure}

GPs are often characterized by having strong EW$_{5007}$ and high SFRs (\citetalias{Cardamone2009}; \citealt{Izotov2011}; \citetalias{Brunker2020}). In Figure \ref{fig:EW_SFR_compare}, we present EW$_{5007,phot}$ plotted against logarithm of the SFR for the SFACT NB659 (red squares), NB695 (orange diamonds), and NB746 (yellow pentagons) GP candidates samples. The gray circles represent the SFACT [\ion{O}{3}] detections that do not satisfy the selection criteria outlined in Table \ref{tab:criteria}. The SFACT spectra do not currently cover a wavelength range that includes the H$\alpha$ emission line, so we derive SFRs of the SFACT GP candidates using the [\ion{O}{3}] $\lambda$5007. However, the use of a fiber-fed instrument like Hydra complicates the measurement of accurate absolute fluxes (\citetalias{Carr2023}), thus making SFRs derived from the Hydra spectra unreliable. Instead, we calculate SFRs using [\ion{O}{3}] luminosities (L$_{[O~III]}$) derived from the flux measured in the NB filter. We correct the [\ion{O}{3}] $\lambda$5007 fluxes for absorption using Balmer reddening coefficients (c$_{H\beta}$) derived following the method presented in \citet{vansistine2016}. The corrected L$_{[O~III]}$ is used to calculate the SFR using the [\ion{O}{3}] star formation calibration presented in \citet{Khostovan2020}. This calibration assumes a Salpeter IMF. The SFRs are presented in Column (10) in Tables \ref{tab: NB659}, \ref{tab: NB695}, and \ref{tab: NB746}. We note that SFR tracers utilizing L$_{[O~III]}$ are sensitive to metallicity, ionization parameter, and nebular excitation which will vary not only across the SFACT GP candidate sample, but across all SFACT [\ion{O}{3}] detections.

We also plot GPs from the \citetalias{Cardamone2009} (blue triangles), \citetalias{Brunker2020} (green X), \citetalias{Kimsey-Miller2024} (purple stars) samples in Figure \ref{fig:EW_SFR_compare}. The values plotted for these comparison samples are spectroscopic EW$_{5007,spec}$ and SFRs calculated using H$\alpha$ emission with the \citet{kennicutt1998} conversion. We note that direct comparisons between [\ion{O}{3}]-based and H$\alpha$-based SFRs is complicated by the selection-dependent biases mentioned. Thus, we restrict our comparison to a qualitative level. The SFACT GP candidates occupy a similarly elevated SFR range as the comparison samples. As previously mentioned, we are in the process of obtaining higher quality spectra that cover a wider wavelength including H$\alpha$. We plan to further explore the SFRs of the SFACT GP candidates when higher quality spectra that include H$\alpha$ are available. 

The range in EW$_{5007,phot}$ decreases with redshift for the SFACT samples, with the NB659 sample covering the same range in EW$_{5007}$ as the \citetalias{Cardamone2009}, \citetalias{Brunker2020}, and \citetalias{Kimsey-Miller2024} GPs for EW$_{5007,phot}$~$\geq$~300~$\text{\AA}$. As mentioned in Section \ref{sec:SFACT_GPs}, we limit our sample to galaxies with EW$_{5007,phot}$~$\geq$ 300~$\text{\AA}$. We note that this is a conservative limit since the comparison samples include galaxies with  EW$_{5007,spec}$~$<$~300~$\text{\AA}$ in their GP catalogs.

The SFACT GP candidates cover a wide range of SFRs, especially the NB659 sample. There is some overlap between the overall SFACT [\ion{O}{3}] sample and the SFACT GP candidate samples in Figure \ref{fig:EW_SFR_compare}. However, none of the overlapping [\ion{O}{3}] detections satisfied all selection criteria. For example, several SFACT [\ion{O}{3}] sources that overlap with the SFACT GP candidates do not satisfy the [\ion{O}{3}]/H$\beta$ criteria. The SFACT GP candidate samples exhibit elevated SFRs relative to the full samples of SFACT [\ion{O}{3}]-detected galaxies in each filter. The NB659, NB695, and NB746 GP candidates samples have median log(SFR) of 0.997, 1.043, and 1.542 M$_{\sun}$~yr$^{-1}$, respectively. These are higher than the median log(SFR) values of the full [O III]-detected samples in each filter by $\sim$0.5 dex in NB659, $\sim$0.2 dex in NB695, and $\sim$0.5 dex in NB746. The elevated SFRs indicate that the GP samples consist of galaxies with enhanced star formation activity. We note the offset in median SFR relative to the full [\ion{O}{3}]-detected samples is slightly lower for the NB695 GP candidates sample than for the NB659 and NB746 GP candidates. This difference likely reflects the inability to apply a color selection criterion for this sample, rather than an intrinsic trend. It is possible that low-mass galaxies with high EW$_{5007,phot}$ but modest SFRs are contaminating the NB695 GP candidate sample, thereby driving down the median SFR of the GP candidates relative to the entire NB695 [\ion{O}{3}] sample.

The relations between stellar mass and SFR or sSFR are important diagnostics for identifying GP-like galaxies. GPs are known to have higher SFRs and sSFRs compared to SFGs with the same stellar mass \citep{Cardamone2009,Izotov2011,Brunker2020,Liu2022,Ding2023,Jeong2025}. Stellar masses are traditionally derived using optical luminosities and colors combined with modeled mass-to-light ratios \citep{BelldeJong2001,Bell2003} or using BB photometry to fit spectral energy distributions (e.g., \citealt{Walcher2011}; \citealt{Hirschauer2018}; \citetalias{Brunker2020}). Star formation activity and, consequently, strong nebular emission present in GPs can cause large discrepancies in the stellar masses derived from mass-to-light prescriptions based on stellar emission. Instead, many previous GP studies use SED fitting with UV, optical, and NIR BB magnitudes to estimate the stellar mass of GPs \citep[\citetalias{Brunker2020};][]{Ding2023,Jeong2025}. However, SFACT is mostly comprised of faint sources (median $g=23.18$; \citetalias{Salzer2023}) that are unlikely to have UV or IR information measured in GALEX \citep{Martin2005} or WISE \citep{Wright2010}. To confirm this, we cross-matched the SFACT GP candidates with the revised GALEX catalog  \citep{Bianchi2017}, the ALLWISE catalog \citep{Cutri2014}, and the unWISE catalog \citep{unWISE2019} using a matching radius of 3\arcsec. Only 36 SFACT GP candidates had a match in one or all of these catalogs, with 9 of them being known H$\alpha$ Dot or KISS GPs. The remaining 27 SFACT GP candidates with GALEX or WISE info have \textit{g}-band apparent magnitudes $g=21.2-23.8$. In many cases, the UV and IR fluxes for these sources carry large uncertainties, making them unsuitable for SED fitting mass analysis.

In the current work, we are unable to reliably determine the stellar masses, and therefore sSFRs, of SFACT detections. Despite this current limitation, using SFACT to catalog a sample of GP candidates yields other advantages. Upon completion of the survey, SFACT will produce a statistically complete galaxy sample useful for a broad range of studies, including redshift evolution studies. As more SFACT fields are completed, we expect the list of GP candidates to grow, allowing us to answer questions about the nature and evolution of this class of galaxies. 
%
\section{Summary and Conclusion} \label{sec:conc}

In this work, we utilize an emission-line selected sample of [\ion{O}{3}]-detected galaxies in the SFACT survey to catalog a new sample of GP candidates in the redshift range z~=~0.31~--~0.51. Our main findings are as follows. 
\begin{enumerate}
    \item For SFACT [\ion{O}{3}] detections in the NB659 and NB746 filters, the [\ion{O}{3}] lines fall in either the ODI \textit{r} or \textit{i} filter, respectively. For NB659 detections, the strong [\ion{O}{3}] boosts the \textit{r} flux, causing high EW$_{5007}$ objects to have bluer $r - i$. Conversely, for NB746 detections, the strong [\ion{O}{3}] boosts the \textit{i} flux, causing high EW$_{5007}$ objects to have redder $r - i$. In these cases, we use color criteria in combination with the photometric EW estimates to identify likely GP candidates. 
    
    \item We are unable to apply a color criterion to identify GP candidates in the NB695 filter. In this redshift range, the [\ion{O}{3}] line shifts from the ODI \textit{r} into \textit{i} bands. As a result, strong [\ion{O}{3}] emitters do not separate from the other star-forming galaxies in $r - i$ in this filter/redshift range. We select GP candidates in the NB695 filter based solely on their high EW$_{5007,phot}$ values. 
    
    \item  We also apply a log([\ion{O}{3}]/H$\beta$) restriction to remove Sy2 and low-excitation sources. We catalog a total sample of 123 candidates comprised of: (1) 64 GP candidates at z~=~0.31~--~0.34; (2) 44 candidates at z~=~0.38~--~0.41; and  (3) 15 GP candidates at z~=~0.48~--~0.51. This is a preliminary list of candidates that is expected to grow as more SFACT data are obtained.
    
    \item The areal density of SFACT GP candidates in the z~=~0.31~--~0.34 window is 2.95~deg$^{-2}$, up to $\sim$40 times higher than in the comparison samples. The increased number of GPs detected in this redshift range reflects the inclusion of many lower-luminosity sources that are more readily observable due to the improved depth of the SFACT survey.

    \item Our GP candidates exhibit elevated SFRs relative to the full [\ion{O}{3}]-detected samples in each redshift window. Their SFRs and EW$_{5007,phot}$ also overlap with GPs previously identified in the SDSS, KISS, and H$\alpha$ Dot surveys, suggesting that the SFACT GP candidates are drawn from the same underlying population.
\end{enumerate}

GPs are in a peculiar evolutionary state, and the GP candidates sample presented in this work provides a robust foundation for investigating the evolution of their physical properties over a broad range of lookback times. Future analyses of this sample will examine their metal abundances and their correlation with star-formation activity, enabling direct comparisons with typical star-forming galaxies and offering new insight into the evolutionary status of GP galaxies.

\begin{acknowledgements}

We would like to gratefully acknowledge the sustained financial support from the College of Arts and Sciences at Indiana University (IU) for the continued operation of the WIYN Observatory. We would also like to express our appreciation for the entire staff of the WIYN observatory for their dedication and hard work in making this research possible. We would like to thank the Indiana Space Grant Consortium Dissertation Fellowship and the IU Department of Astronomy Sullivan Fellowship for partial funding of this research. K.N.B. would like to thank Dr. Katherine Rhode for her helpful discussion on redshift tracks. We would also like to thank the anonymous reviewer who provided insightful suggestions for improving the manuscript. 

\facilities{WIYN (ODI: \citealt{Harbeck2010}; Hydra: \citealt{Barden1994,Barden1995})}

\software{\texttt{astropy}  \citep{astropy:2013,astropy:2018,astropy:2022}, CosmologyCalculator \citep{Wright_2006}, \texttt{matplotlib} \citep{Hunter:2007}, \texttt{numpy} \citep{numpy}, \texttt{pandas} \citep{pandas_17229934}, \texttt{scipy} \citep{2020SciPy-NMeth,scipy_17467817}}

\end{acknowledgements}


\bibliography{Baker_SFACT_GP_revised_v2}{}
\bibliographystyle{aasjournalv7}



\end{document}